\documentclass[12pt]{iopart}

\usepackage{enumerate}
\usepackage{isotope}
\usepackage{lipsum}
\usepackage{bbm}
\usepackage{graphicx}
\usepackage{feynmp}
\usepackage{multirow,dcolumn}
\usepackage{amssymb}
\usepackage{bm}
\usepackage{amsfonts}
\usepackage{stackrel}
\usepackage[normalem]{ulem}

\newcolumntype{.}{D{.}{.}{1.5}}
\newcolumntype{,}{D{.}{.}{1.8}}
\newcolumntype{;}{D{.}{.}{2.3}}
\newcolumntype{+}{D{.}{.}{5.3}}
\newcolumntype{-}{D{.}{.}{3.4}}

\usepackage{iopams}  
\usepackage{hyperref}
\usepackage[all]{hypcap}
\hypersetup{colorlinks=true,citecolor=[rgb]{0,0,0.8},
linkcolor=[rgb]{0,0,0.8},urlcolor=[rgb]{0,0,0.8}}

\usepackage[usenames]{color}
\definecolor{jlred}{rgb}{0.81,0.13,0.16}

\definecolor{ingocolor}{rgb}{0.13,0.81,0.16}

\definecolor{aekcolor}{rgb}{0.13,0.16,0.81}

\begin{document}

\topical[New Ideas in Constraining Nuclear Forces]{New Ideas in Constraining Nuclear Forces}

\author{Ingo Tews$^{1}$, Zohreh Davoudi$^{2,3}$, Andreas Ekstr\"om$^4$, Jason D. Holt$^5$, and Joel E. Lynn$^{6,7}$,}

\address{$^1$ Theoretical Division, Los Alamos National Laboratory,
Los Alamos, New Mexico 87545, USA}
\address{$^2$ Maryland Center for Fundamental Physics and Department of Physics, University of Maryland, College Park, Maryland 20742, USA}
\address{$^3$ RIKEN Center for Accelerator-based Sciences, Wako 351-0198, Japan}
\address{$^4$ Department of Physics, Chalmers University of Technology, SE-412 96 G\"{o}teborg, Sweden}
\address{$^5$ TRIUMF, 4004 Wesbrook Mall, Vancouver, BC V6T 2A3, Canada}
\address{$^6$ Institut f\"ur Kernphysik,
Technische Universit\"at Darmstadt, 64289 Darmstadt, Germany}
\address{$^7$ ExtreMe Matter Institute EMMI, GSI Helmholtzzentrum f\"ur
Schwerionenforschung GmbH, 64291 Darmstadt, Germany}

\ead{itews@lanl.gov}
\vspace{10pt}
\begin{indented}
\item[]August 2020
\end{indented}

\begin{abstract} 
In recent years, nuclear physics
has benefited greatly from the development of powerful \textit{ab
initio} many-body methods and their combination with interactions from
chiral effective field theory. With increasing computational power and
continuous development of these methods, we are entering an era of
precision nuclear physics. Indeed, uncertainties from nuclear Hamiltonians
now dominate over uncertainties from many-body methods. This
review summarizes the current status of, and future directions
in, deriving and constraining nuclear Hamiltonians.
\end{abstract}

%
\vspace{2pc}
\noindent{\it Keywords}: nuclear interactions, chiral effective field theory, many-body methods, Lattice QCD
%
%
%
%

\section{Introduction}

A precise and accurate description of the interactions between protons
and neutrons (nucleons) is the key component in understanding many
important physical observables over a wide range of length and energy
scales, from neutrinos to neutron stars. From fundamental symmetries of the micro-cosmos to the stellar explosions in the macro-cosmos where the elements are formed, nuclear forces play a central role in research at the forefront of nuclear physics. Some relevant examples include the possible existence of few-neutron
resonances~\cite{Kisamori:2016jie,Gandolfi:2016bth,Fossez:2016dch},
nuclear structure observables in the medium- to heavy-mass region of
the nuclear chart~\cite{Hergert:2015awm,hagen2016}, which set the
nucleosynthesis path far away from stability, and astrophysical
phenomena, such as properties of neutron
stars~\cite{hebeler2015}.
A systematically improvable and well-founded theory for the nuclear Hamiltonian is of
major relevance to reliably answer these and many other scientific questions.

\begin{figure}[t]
\centering
\includegraphics[trim=11.0cm 4.5cm 2.0cm 4cm,clip=,width=0.6\columnwidth]{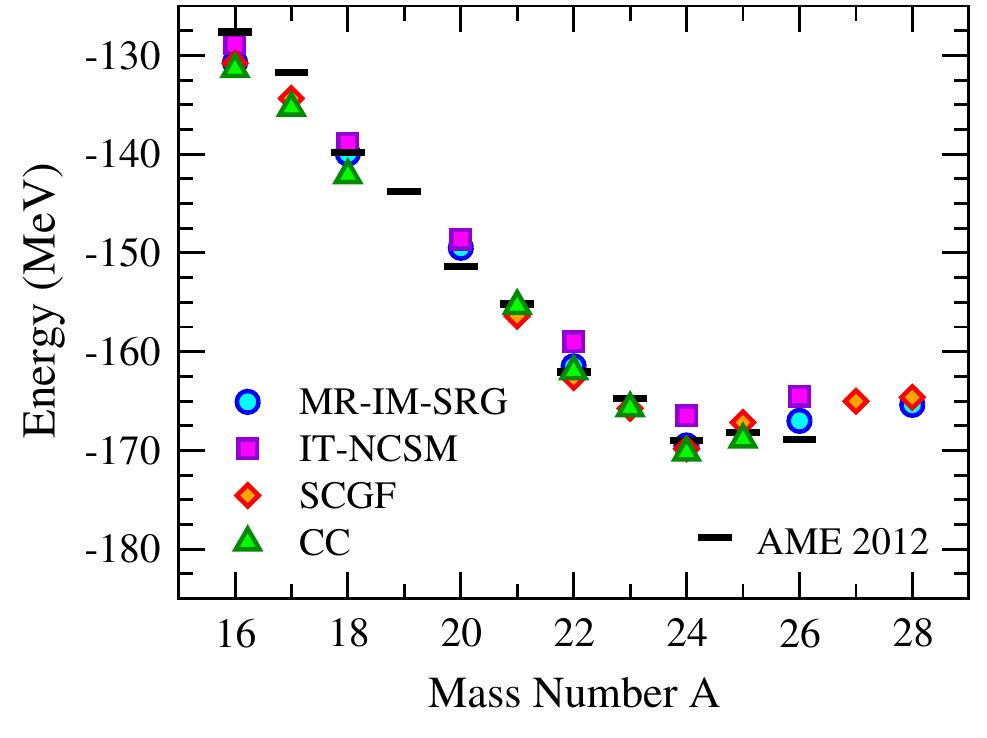}
\caption{\label{fig:MBcomparison}
Comparison of ground-state energies in the oxygen isotopic chain from
different many-body methods. 
Figure reprinted with permission from Ref.~\cite{hebeler2015}.}
\end{figure}

Within the last few decades, significant progress in theoretical nuclear physics
has been made possible, mainly due to the theoretical and algorithmic
development of powerful \textit{ab initio} many-body methods and their
combination with renormalization group (RG)
techniques~\cite{PhysRevC.75.061001,BOGNER20031} and effective field
theories (EFT)~\cite{Weinberg1979, Weinberg:1991um, Bedaque:2002gm, Epelbaum:2008ga, Machleidt:2011zz}. Different many-body methods
employ different mathematical approaches to approximately solve the
the same many-body Schr\"odinger equation and some of the most common approaches are 
quantum Monte Carlo (QMC) methods~\cite{Carlson:2014vla}, the no-core
shell model (NCSM)~\cite{Barrett:2013nh}, the coupled-cluster (CC)
method~\cite{Hagen:2013nca,Hagen:2015yea}, the self-consistent Green's
function (SCGF) method~\cite{Carbone:2013eqa,Soma:2013xha}, the
in-medium similarity renormalization group (IMSRG)
method~\cite{Hergert:2015awm,Stro19ARNPS}, and nuclear lattice
methods~\cite{Epelbaum:2011md}. With increasing computational power and continuous development of available many-body methods, microscopic calculations of nuclear systems become increasingly precise and uncertainties stemming from the many-body methods themselves can be controlled and accounted for in a systematic way. In addition, there is remarkable agreement (see, e.g., Ref.~\cite{hebeler2015}
and Fig.~\ref{fig:MBcomparison}) between the results from different many-body
approaches~\cite{Gandolfi:2015jma,hebeler2015}. This consistency, in
combination with the attained numerical convergence, suggests that the
method uncertainties induced by different mathematical approximations in the solution of the many-body Schr\"odinger equation
are well-controlled and, most importantly, that
uncertainties from the nuclear Hamiltonian, i.e. the uncertain
theoretical description of the nuclear potential, now
dominate over uncertainties from many-body methods. This development
is reflected in the recent emergence of a plethora of different
nuclear interaction models. Indeed, 15 years ago, when \textit{ab
  initio} methods struggled to converge in $p$-shell nuclei, and most
efforts focused on pushing the frontier of numerical many-body
calculations and the inclusion of three-nucleon interactions, only a handful of different nuclear interaction models
were routinely employed; most prominently the
Idaho-N3LO(500)~\cite{Entem:2003ft}, CD-Bonn~\cite{Machleidt:2000ge},
and AV18~\cite{Wiringa:1994wb} two-nucleon ($NN$) interactions.

Phenomenological
approaches to the description of the nuclear interaction, such as the CD-Bonn and AV18 $NN$ potentials, are primarily guided by the quantitative reproduction of
experimental data.
In contrast, most present-day efforts aim at constructing
effective field theories (EFTs) of quantum chromodynamics (QCD), the fundamental theory of the strong interactions between quarks and gluons.  

An EFT is a versatile tool in theoretical physics, and any EFT description of some underlying theory, in our case QCD, requires a separation of scales: a separation of a hard scale, $\Lambda_b$, in the system under study that is larger than the typical external momentum scale $p$, or soft scale, of the processes one would like to study. 
In principle, such an EFT can be constructed in a top-down approach.
After identifying scale separations, the next step in constructing an EFT of a 
more fundamental quantum field theory is to integrate out the high-momentum contributions
($p>\Lambda_b$) in the quantum-mechanical path integral. 
This leads to a non-local theory. An operator product expansion in local operators is performed~\cite{Collins:1984xc} whose coefficients encode short-distance physics and can be identified with so-called low-energy constants (LECs). The LECs contain all of the information about
the high-energy theory (i.e., QCD in nuclear physics) that is relevant for computing low-energy observables. 
The resulting effective Lagrangian describes the EFT with factorized long- and
short-distance physics.
Up to this point, no approximations have been made and the original information content of the theory is preserved.

In practice, however, this top-down approach is difficult to implement. Instead, the EFT is constructed in a bottom-up approach.
The starting point for constructing a nuclear EFT is an effective Lagrangian consisting of all possible terms that are consistent with the known symmetries and symmetry-breaking patterns of QCD, e.g., 
Lorentz invariance, color $SU(3)$ gauge symmetry, electromagnetic $U(1)$, and
the approximate and spontaneously broken chiral symmetry $SU(2)_{L}
\times SU(2)_{R}$, which also leads to the emergence of pions as
pseudo-Nambu-Goldstone bosons. Depending on the EFT, the effective Lagrangian is built from pions, nucleons, and sometimes the $\Delta$ resonance, and, in principle, contains an infinite number of terms. 
Therefore, also an infinite number of LECs with \textit{a priori} unknown numerical values need to be determined. 
The upshot of the EFT approach lies in an approximation whereby the observable is expanded in the small ratio $k/\Lambda_b \ll 1$, where $k$ is the soft scale of the problem, e.g., $k=\{m_{\pi},p\}$ with the pion mass $m_{\pi}$ in the case of chiral EFT. This expansion is then truncated at some sufficiently
low and finite order, but including higher orders
should increase the precision of the final result. The connection
between the $k/\Lambda_b$-expansion and the set of effective interactions that must be retained at each order in the expansion is called the \textit{power
  counting} of the EFT. A consistent
power counting (PC) ensures that physical observables are independent
of the choice of ultraviolet cutoff employed in the regularization scheme, which is necessary when implementing nuclear interactions in many-body methods to remove high-momentum contributions that otherwise would lead to divergences. 
The systematic expansion of the EFT promises a handle on the systematic uncertainty in theoretical predictions of nuclear observables.
The possibility of linking nuclei with QCD in a systematic fashion makes chiral EFT a highly attractive framework.

The choice of $\Lambda_b$ confines the choice of the low-energy degrees
of freedom. For instance in QCD, with a hard scale $\Lambda_b = m_{\pi}$, pionic degrees of freedom are integrated out. The
resulting pionless EFT~\cite{VANKOLCK1999273,Chen:1999tn} offers several
advantageous avenues for analytical work and studies of
renormalizability, but its domain of applicability beyond the lightest
nuclei is unknown and currently being
explored~\cite{platter2005,Barnea:2013uqa,Contessi:2017rww,Bansal:2017pwn}. Once 
the relevant soft scale approaches $m_{\pi}$, the hard scale must be shifted. A natural choice is the spontaneous chiral symmetry breaking scale, roughly the mass of the $\rho$ meson,
$\Lambda_b \sim m_{\rho}$.
Consequently the EFT must explicitly include the pion. This formulation is called
pionfull EFT, or chiral EFT~\cite{ORDONEZ1992459,Ordonez:1993tn}. In
contrast to pionless EFT, observables in chiral EFT can be challenging to obtain analytically
even in the few-nucleon sector, and predictions rely on numerical studies. In addition, nucleon resonances might have to be explicitly included in the effective Lagrangian, e.g., the $\Delta$ isobar when $p\sim 2m_{\pi}$~\cite{vanKolck1994}.

Chiral EFT is a powerful tool for nuclear physics but there are many open problems that temper its promises. First, and foremost, the choice of power
counting in chiral EFT is an unsettled
matter~\cite{Epelbaum:2006pt,Phillips12,Griesshammer:2015osb,Nogga:2005hy,valderrama2016}. It should be
pointed out that there is no guarantee that a consistent power
counting must exist. Second, to determine the numerical values of the finite number of LECs up to
a given order in the $k/\Lambda_b$ expansion, i.e., to calibrate the theory to data, typically leads to a
nontrivial parameter estimation or optimization problems~\cite{Ekstrom:2013kea,Ekstrom:2015rta,Carlsson:2015vda,Melendez:2017phj,Wesolowski:2018lzj}. Although not yet feasible, a direct Monte Carlo evaluation of the QCD path integral on
a space-time lattice using lattice QCD methods could provide
theoretical input for the LECs and/or valuable
pseudo-data~\cite{Chang:2018uf}. Third, quantifying the probability distribution of the systematic EFT LECs in predictions of important nuclear observables requires statistical inference procedures that are currently being developed~\cite{Furnstahl:2015rha,Melendez:2017phj,Carlsson:2015vda}. Fourth, and last, the atomic nucleus exhibits a
multitude of characteristic scales (shallow bound states, spatially
extended halo states, cluster states, collective states, etc.)
and it is not clear at the outset which is the optimal type of EFT for
a particular problem. Linking these different EFTs for atomic nuclei deserves much more attention. 
Also, different many-body methods specialize in different classes of observables, e.g., deformed vs. spherical states, single-particle vs. collective excitations, clusterization, etc. Therefore, methods for transferring information about EFT interaction models obtained using different many-body methods to other methods could also be important. 

These open questions lead to sizable uncertainties in the nuclear
Hamiltonian, which in turn affect several observables that are within the reach of
\textit{ab initio} many-body methods, including neutron matter and
saturation properties of nuclear
matter~\cite{Drischler:2015eba,Lynn:2015jua}, properties of nuclei up to the medium-mass
region~\cite{Hagen:2015yea}, currents~\cite{Klos:2016omi} and
reactions~\cite{Lynn:2015jua,Elhatisari:2015iga,Rotureau:2016jpf,Calci:2016dfb}.
Currently, the uncertainties of the nuclear Hamiltonian are the main
limitation in accurate predictions of these and related systems, such
as the equation of state of nuclear matter~\cite{Tews:2018kmu}, the
nuclear symmetry energy and its density dependence~\cite{Lim:2019som}, the location of the proton and neutron driplines~\cite{Holt19Drip}, nuclear
energy levels and the evolution of magic numbers far from stability~\cite{Stroberg:2015ymf,Liu:2018mtu,Tani1978Ni}, or \textit{ab initio}
nuclear matrix elements for single $\beta$~\cite{Gysbers:2019uyb} and neutrinoless double $\beta$ decays~\cite{Engel_2017}.
These limitations further impact predictions for astrophysics, e.g.,
for the structure and the mass-radius relation of neutron stars,
nucleosynthesis and the r-process path, and simulations of supernovae
and neutron-star mergers.  To improve the nuclear Hamiltonian,
significant effort has been invested within the last few years in
several complementary directions.  Besides developing optimization and parameter-estimation strategies,
it is important to examine different regularization schemes~\cite{Entem:2003ft, EGMN3LO, Navratil:2007zn, Gezerlis:2013ipa, Epelbaum:2014efa}, and explore paths of matching nuclear interactions to lattice QCD
simulations~\cite{Contessi:2017rww}.

To address the successes and shortcomings of current microscopic approaches to analyze and predict nuclear systems,  the authors of this review organized, and/or participated in, a
workshop at the European Centre for Theoretical Studies in Nuclear
Physics and Related Areas (ECT*) in June 2018~\cite{Workshop}, bringing
together nuclear physicists that currently work in the
above-mentioned directions, i.e., specialists in the ``construction''
of nuclear Hamiltonians and experts in many-body calculations of
systems ranging from nuclei to nuclear matter, with the goal of
identifying possible future pathways and novel constraints that can be
used to further improve our understanding of nuclear interactions.
Given the diversity of the participants' research areas and the
opposing viewpoints on some of the topics, the workshop
was very lively and filled with several long debates and fruitful discussions.
In this review, we present some of the problems that were
discussed during the workshop and also present conclusions drawn by the
participants. Though consensus was not always reached, the discussions and presentations clarified some open questions, spawned
new research projects, coordinated some of the efforts within
the community, and contributed towards a common and organized approach to
further improve nuclear interactions. All of the above are necessary for enabling future
high-precision calculations of nuclei and nuclear matter for applications in nuclear physics
and nuclear astrophysics.

\section{Current Nuclear Hamiltonians}

Realistic predictions of nuclear observables require accurate many-body  methods and reliable nuclear Hamiltonians with quantified theoretical uncertainties. Typically, the nuclear Hamiltonian
can be written as
\begin{equation}
H=T+V_{NN}+V_{3N}+V_{AN}\,,
\end{equation}
where $T$ is the kinetic energy of the nucleons, $V_{NN}$ is the $NN$ interaction, $V_{3N}$ is the three-nucleon ($3N$)
interaction, and $V_{AN}$ contains all higher-nucleon forces. Given current theoretical uncertainties, it is
usually sufficient to truncate the Hamiltonian at the $3N$ level, except if a high-precision is sought.
This was shown explicitly, e.g., in nucleonic
matter~\cite{Kruger:2013kua}. In addition, based on PC arguments, four- and higher-nucleon forces are expected to be
small in presently-used chiral EFT formulations. 
However, the validity if this truncation needs to be explicitly checked in future calculations.
For many years, the
Hamiltonian was constructed phenomenologically, i.e., without a fundamental guiding principle,
by modeling the
nuclear interactions as, e.g., a sum of boson-exchange
potentials~\cite{Machleidt:2000ge}, or by constructing Hamiltonians
with a given spin-isospin operator structure and radial functions~\cite{Wiringa:1994wb}. These were fit
to reproduce $NN$ scattering data with a small $\chi^2$ up to
relatively high scattering energies (see, e.g.,
Ref.~\cite{Wiringa:1994wb}). In addition, such $NN$ forces were
supplemented with phenomenological $3N$ forces (e.g. the Urbana
IX~\cite{Pudliner:1997ck} and Illinois~\cite{Pieper:2001ap} $3N$
potentials) with great success~\cite{Wiringa:2002ja}. These many-body
forces, however, were not constructed consistently with the $NN$
sector. Furthermore, while the Urbana IX $3N$ interaction led to a
reasonable description of neutron matter~\cite{akmal:1998cf}, it failed
to produce adequate spin-orbit splitting in light
nuclei~\cite{Pieper:2001ap}.  The Illinois-7 $3N$ potential, on the other hand, 
led to an excellent description of nuclei~\cite{Carlson:2014vla},
but failed in neutron matter, producing a vanishing pressure above a density
of approximately 0.1~$\mathrm{fm}^{-3}$~\cite{Maris:2013rgq}.

\begin{figure}[t]
\centering
\includegraphics[trim= 0.0cm 1cm 0cm 1cm, clip=,width=0.95\columnwidth]{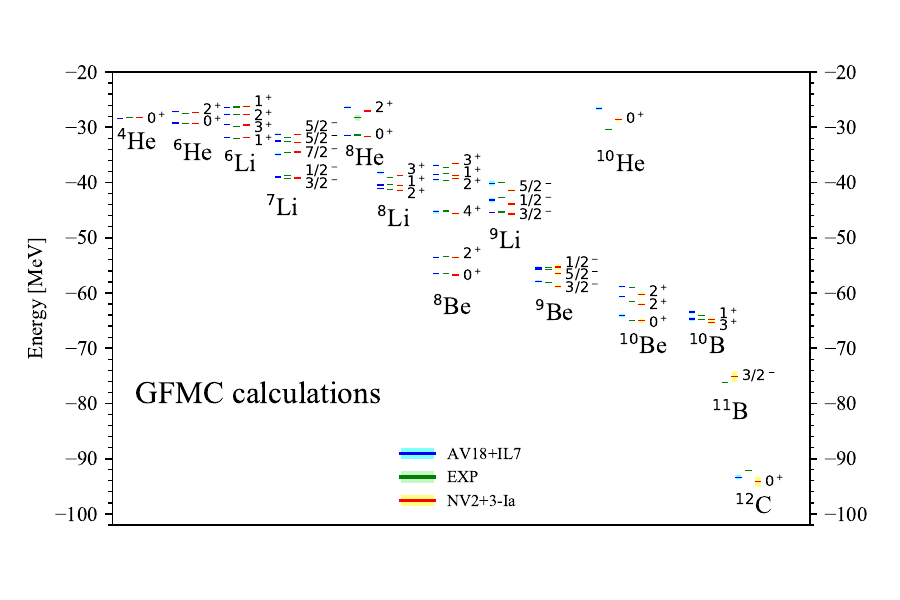}
\caption{\label{fig:compGFMC} Energy spectra for light nuclei up to $^{12}$C for GFMC calculations using the AV18+IL7 phenomenological interactions~\cite{Wiringa:1994wb,Pieper:2001ap} or local $\Delta$-full chiral interactions~\cite{Piarulli:2017dwd}. Reprinted figure with permission from Ref.~\cite{Piarulli:2017dwd}. Copyright 2017 by the American Physical Society.}
\end{figure} 

Many of these shortcomings were remedied with the advent of chiral EFT. In particular, chiral EFT establishes a link between $NN$ and multi-nucleon forces, as well as low-energy currents. The EFT expansion has a lowest order, the so-called leading order (LO), $(k/\Lambda_b)^0$, and is typically truncated at some sub-leading order, e.g. at third (N$^2$LO), fourth (N$^3$LO), or even fifth (N$^4$LO) order. Recent findings~\cite{Furnstahl:2015rha,Melendez:2017phj} indicate that the hard scale $\Lambda_b$, or breakdown scale, of chiral EFT is approximately of the order of 500-600~MeV.
The exact form of the expansion is given by a PC scheme which we will discuss in more detail in Sec.~\ref{sec:PC}.
The resulting truncated interaction is then included in a given many-body method. 
Chiral Hamiltonians present several advantages over older
phenomenological interactions. They provide: (i) a systematic scheme to
improve the nuclear Hamiltonian, (ii) consistent $NN$,
$3N$, and $AN$ interactions, and (iii) means of assessing
the theoretical uncertainty. In addition, chiral Hamiltonians have reached a similar precision to phenomenological interactions, see, e.g., Fig.~\ref{fig:compGFMC} for a comparison of phenomenological and chiral interactions in GFMC calculations of light nuclei up to $^{12}$C.

At this point, we give an overview of commonly used chiral potentials. Chiral interactions can be separated according to their regularization scheme into nonlocal, semilocal, and local interactions. These schemes will be described in the following sections. The first two classes are commonly given in momentum space, while local potentials are usually developed in coordinate space. 
The first generation of chiral interactions that were extensively used in \textit{ab initio} calculations were nonlocal N$^3$LO interactions that were developed in the early 2000s~\cite{Entem:2003ft,EGMN3LO}. Recently, many new nonlocal $NN+3N$ interactions have been developed at N$^2$LO~\cite{Ekstrom:2013kea,Carlsson:2015vda,Ekstrom:2015rta}, using mathematical optimization algorithms. In addition, nonlocal $\Delta$-full interactions, i.e. interactions with explicit inclusion of the $\Delta$-isobar degree of freedom, have also been constructed up to this order~\cite{Ekstrom:2017koy}. There also exist nonlocal $NN$ interactions at N$^4$LO~\cite{Entem:2017gor}. 
In the semilocal sector, where short-range physics is regulated nonlocally and long-range physics locally, interactions have been developed up to N$^4$LO~\cite{Reinert:2017usi}. 
Finally, local interactions have been developed specifically to be employed in quantum Monte Carlo methods, and include the delta-less N$^2$LO interactions developed in Refs.~\cite{Gezerlis:2013ipa,Gezerlis:2014zia,Tews:2015ufa,Lynn:2015jua}, and the $\Delta$-full Norfolk interactions of Refs.~\cite{Piarulli:2014bda,Piarulli:2017dwd}.

Even though these modern interactions have been used very successfully
in calculations of selected nuclear systems, they currently represent the major source of uncertainty in calculations of nuclear systems.
These uncertainties arise from the truncation of the chiral
expansion at a certain order, the uncertainty of the two- and many-body LECs, and the regularization scheme and scale dependence.
However, there are many developments to remedy these limitations, as presented in detail in Sec.~\ref{sec:limitations}.
These include constraining the interactions with different observables (Sec.~\ref{sec:observables}), uncertainty quantification, optimization, or inclusion of higher orders in the EFT (Sec.~\ref{sec:fitting}), new developments in PC schemes (Sec.~\ref{sec:PC}), and matching interactions to lattice QCD calculations (Sec.~\ref{sec:lqcd}).

\section{Current Limitations of Nuclear Hamiltonians}~\label{sec:limitations}

\noindent
Nuclear Hamiltonians suffer from
three major sources of uncertainty: the truncation of the chiral
expansion which results in a description of physical systems with an insufficient operator basis, the related uncertainty of LECs, which additionally originates from
fits to a poor database or fits to nonideal systems, and the dependence on
the regularization scheme and scale. Before presenting possible
remedies in the next sections, let us explain these issues in more detail.

\subsection{Truncation Uncertainty}

Chiral EFT is a low-momentum expansion in terms of a
dimensionless expansion parameter $Q = \{m_{\pi},p\}/\Lambda_b$, that contains
information on the momentum scale $p$ relevant for the process under study, as well as
the breakdown scale of the EFT. Following the discussion of,
e.g., Ref.~\cite{Furnstahl:2015rha}, one can express an observable $X$
as a sum of all interaction terms:
\begin{equation}\label{eq:EFTexpansion}
X=X_0\sum_{k=0}^{\infty}c_k Q^{k}\,,
\end{equation}
where $X_0$ sets the scale expected for the observable $X$ and
can be chosen as, e.g., the leading-order result, $c_0 X_0=X_{\rm{LO}}$. Since
the EFT expansion is truncated, i.e., the above sum is cut off at a
certain $k=k_{\rm max}$, the Hamiltonian necessarily has an intrinsic uncertainty due to the neglected terms at $k>k_{\rm max}$,
which should account for the missing interaction terms. The highest power
$k_{\rm max}$ of $Q$ in the chiral Hamiltonian defines the order at which one is
working. The uncertainty $\Delta X$ is given by
\begin{equation}
\Delta X=X- X_0\sum_{k=0}^{k_{\rm max}}c_k Q^{k}=X_0\sum_{k=k_{\rm max}+1}^{\infty}c_k Q^{k}\,.
\end{equation}

As a guiding uncertainty estimate, if $Q$ is not too large, it is usually sufficient to estimate the magnitude of the first omitted term $X_0 c_{k_{\rm max}+1} Q^{k_{\rm max}+1}$ where the magnitude of the unknown expansion coefficient can be estimated as the absolute maximum of the $k \leq k_{\rm max}$ coefficients. 
Estimating the coefficient in that way prohibits $c_{k_{\rm max}+1}$ to be unnaturally small compared to previous coefficients. 
Thus, the higher the order at which one is working, the more information is provided to estimate the uncertainty, and the higher is the degree of belief in
the resulting uncertainty band. This uncertainty is similar to the one
presented by Epelbaum, Krebs, and Mei{\ss}ner (EKM)~\cite{Epelbaum:2014efa}.

These truncation uncertainties are inherent in any nuclear EFT interaction, and, hence, should always be estimated. Otherwise, when uncertainty estimates are not provided, comparisons between different nuclear Hamiltonians will not be meaningful, and no firm conclusion about the quality of predictions from individual Hamiltonians can be drawn. In addition, in the context of EFTs, a determination of truncation uncertainties is necessary to assess the convergence of the expansion, which is not clear at the moment in certain systems. 

\begin{figure}[t]
\centering
\includegraphics[trim= 0 0 0 0, clip=,width=1.0\columnwidth]{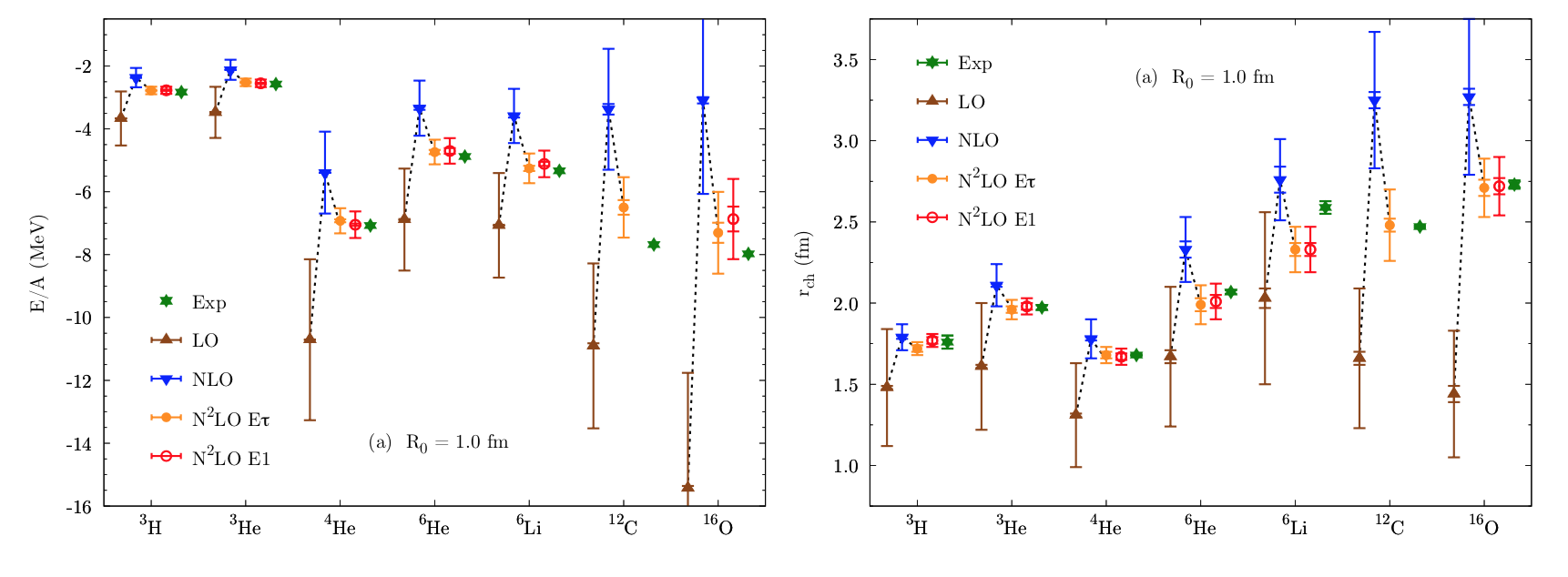}
\caption{\label{fig:EKMnuclei}
Quantum Monte Carlo results for ground-state energies and charge radii of selected nuclei up to \isotope[16]{O} with EKM uncertainties. Reprinted figures from Refs.~\cite{Lonardoni:2017hgs,Lonardoni:2018nob}. Copyright 2018 by the American Physical Society.}
\end{figure} 

The above prescription for estimating the theoretical uncertainties has been applied to several calculations of atomic nuclei and neutron matter. Figure~\ref{fig:EKMnuclei}, for example, presents the results of QMC calculations from Refs.~\cite{Lonardoni:2017hgs,Lonardoni:2018nob} carried out at different chiral orders. 
At N$^2$LO, several sources of uncertainty are indicated. The difference between orange and red uncertainty bars depends on the regularization scheme, which in this case is small at this particular cutoff scale. We will discuss this issue in more detail later. The smaller uncertainty bars indicate the QMC statistical many-body method uncertainties, while the larger uncertainty bars originate from the truncation of the chiral expansion. It is evident that truncation uncertainties are dominating.

While the discussed prescription to estimate uncertainties is rather simple and very useful, several problems remain that have to be addressed in order to clarify existing challenges with the chiral EFT expansion. 
As pointed out earlier, the series expansion coefficients $c_i$ in Eq.~(\ref{eq:EFTexpansion}) can be used to analyze the convergence of the chiral expansion, given that the expansion parameter $Q$ is known~\cite{Furnstahl:2015rha}. 
The coefficients in the expansion can be used to estimate the breakdown scale $\Lambda_b$ if the typical momentum scale of the system is known, and vice versa. 
However, these questions are far from trivial. For instance, the relevant momentum scale in a bound many-body state, or for nuclear matter at a certain density, is difficult to quantify \textit{a priori}. 
While determinations of the breakdown scale using Bayesian analysis of $NN$ scattering data put it around 600 MeV~\cite{Furnstahl:2015rha, Melendez:2017phj} for the semilocal interactions of Refs.~\cite{Epelbaum:2014efa,Epelbaum:2014sza}, the question is if this scale persists also in a many-body system. 
These problems make it difficult to judge the convergence of the chiral series in scenarios involving many nucleons. 
This question is especially relevant for astrophysical applications, e.g., in nuclear matter probed in neutron stars, where densities beyond typical nuclear densities are explored. In particular, the density range between $1-2 n_{\rm{sat}}$ is crucial for neutron-star physics. 
For these applications, it remains unclear up to which densities chiral EFT interactions remain reliable.

\subsection{Scheme and Scale Dependence and PC}

Intimately connected to the first issue, i.e. the truncation uncertainty and EFT convergence, is the question of scheme and scale dependence. 
To converge in many-body calculations, it is necessary to apply a regularization scheme to EFT interactions that cuts off, in a certain way, momenta above a certain scale (referred to as the cutoff). 
In a well-defined EFT, observables should be independent of the chosen scheme at sufficiently large cutoffs. However, there are clear indications that current nuclear Hamiltonians based on canonical chiral EFTs lead to results that depend both on the chosen regularization scheme as well as the value of the cutoff and, hence, suffer from regulator artifacts. 
For instance, it has been found that local regulators lead to very different regulator artifacts than nonlocal regulators, see, e.g., Refs.~\cite{Tews:2015ufa,Dyhdalo:2016ygz, Huth:2017wzw}. 

These regulator artifacts in general depend on $(p/\Lambda)^{n}$, where
$\Lambda$ is the cutoff. If $\Lambda$ is chosen sufficiently high, then the regulator artifacts become small and uncertainties are dominated by the breakdown scale. 
However, current nuclear Hamiltonians are restricted to a rather limited range of possible cutoff values due to shortcomings in the employed PC scheme and the appearance of spurious bound states. In particular, current chiral EFT Hamiltonians are based on Weinberg PC, see Sec.~\ref{sec:PC}, which is a power counting scheme in terms of the potential as opposed to the physical amplitude. 
A chiral EFT based on Weinberg PC is not renormalizable due to an infinite-order iteration of the finite-order EFT potential in the solution of the Schr\"odinger equation. 
The results are then dependent on the cutoff value even at very high values of the cutoff.  For example,
increasing the cutoff leads to the appearance of spurious bound states that may be a problem for certain many-body methods. In addition, 
high-cutoff interactions have a practical drawback because they are usually harder and do not converge sufficiently fast in many-body interactions. There is a significant effort
to construct improved regularization schemes, e.g., semilocal
regulators~\cite{Reinert:2017usi}, or high-cutoff interactions~\cite{Nogga:2005hy,Tews:2018sbi}, but many questions remain open, including how to regularize currents in such frameworks.

Consequently, the issue of scheme and scale dependence is deeply related to the construction and definition of a consistent PC, if it exists at all. This leads to the question of whether the chiral expansion is converging within Weinberg PC. 
For instance, Weinberg PC is based on naive dimensional analysis of the interactions and grounded in the hope that nuclear observables experience the same convergence behavior as the interaction potential. While this may be an \textit{a priori} reasonable expectation, it is not clear if many-body systems behave this way. 
For instance, combinatorial factors in large-$A$ systems might change this picture and lead to a breakdown of Weinberg PC in large many-body systems. 

Questions about the convergence of chiral EFT, its power counting, and related issues have to be addressed by the whole community. 
Currently, this presents a serious limitation to the viability of using chiral EFT interactions to connect many-body systems to the underlying theory of strong interactions. We will come back to the PC issue in Sec.~\ref{sec:PC}. 
 We turn to the issue of LEC uncertainties next, and 
present some examples on the convergence of chiral EFT in the next sections. 

\subsection{Uncertainty in the Values of LECs}

All of the LECs in nuclear Hamiltonians carry an uncertainty that originates from limited datasets or experimental uncertainties for fit observables. 
This uncertainty is especially large in the $3N$
sector. For the leading-order $3N$ forces, five LECs need to be determined. 
All but two of them, denoted $c_i$ in the literature, also appear in the subleading two-pion--exchange interaction in the $NN$ sector, and their uncertainty can be partially absorbed by other $NN$ LECs. In the $3N$ sector, they determine the strength of the $3N$ two-pion exchange and can have a sizable impact on the results, e.g., in nuclear matter~\cite{Hebeler:2009iv}. 
The convergence of pion-nucleon scattering observables and the values of the $c_i$ LECs were studied using the Roy-Steiner formalism~\cite{Hoferichter:2015tha, Hoferichter:2015hva} which will be discussed in the next Section. In this context, it is clear that the inclusion of the $\Delta$ degree of freedom improves the convergence of these observables and leads to LECs with a ``natural'' size of $\mathcal{O}(1)$.

\begin{figure}[t]
\centering
\includegraphics[trim= 0 1.5cm 0 2.0cm, clip=,width=1.0\columnwidth]{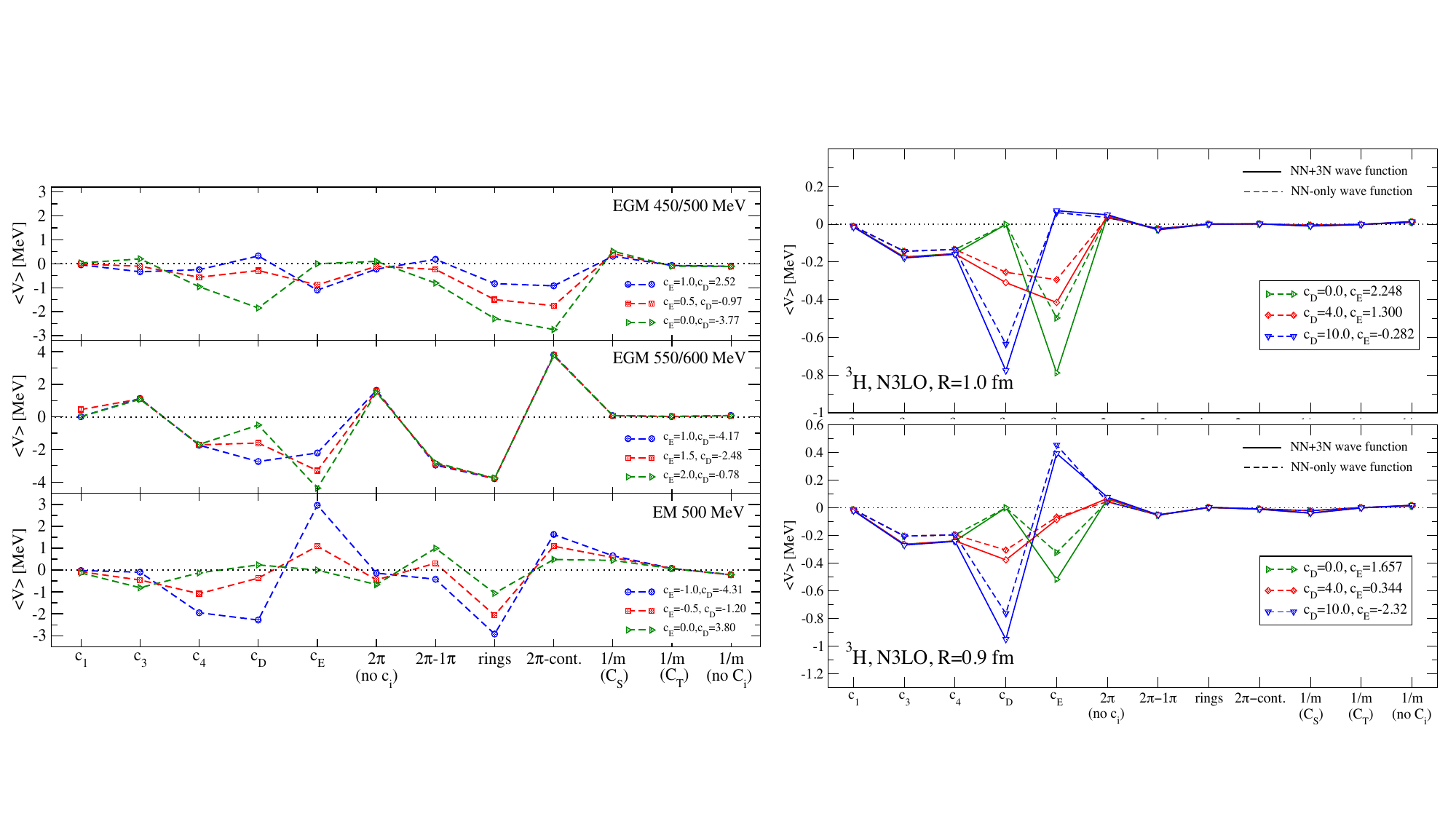}\\
\includegraphics[trim= 0 0 0 0, clip=,width=1.0\columnwidth]{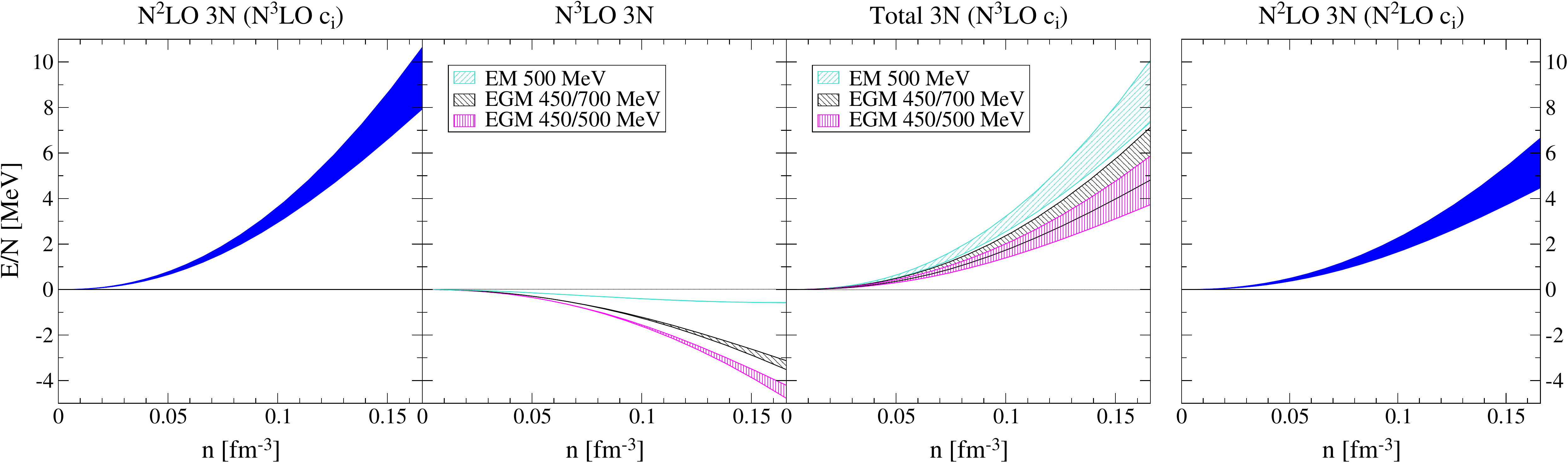}
\caption{\label{fig:3Ncomparison} 
[Upper panels] The convergence of individual $3N$ contributions in \isotope[3]{H} for nonlocal (left) and semilocal (right) regulators. In each panel the first five points denote N$^2$LO 3N topologies, while the last seven points denote N$^3$LO topologies.
It can be seen that some N$^3$LO $3N$ topologies might be of similar size as N$^2$LO contributions. 
[Lower panels] The convergence of the total
  $3N$ contribution in neutron matter for nonlocal regulators at the Hartree-Fock level. The left two panels show the leading and
  subleading $3N$ contributions at N$^3$LO and the right two panels
  show the total $3N$ contributions at N$^2$LO and N$^3$LO. Reprinted figures with permission from Refs.~\cite{Tews:2012fj, Hebeler:2015wxa}.
  Copyright 2013 and 2015 by the American Physical Society.}
\end{figure} 

It has also been found that individual
subleading $3N$ topologies can be comparable in size to leading $3N$
topologies and experience large cancellations, depending on the regularization scheme~\cite{Hebeler:2015wxa}, see
the upper panels of Fig.~\ref{fig:3Ncomparison} for nonlocal and semilocal
regulators. This may be a signature of an inconsistent PC or a regulator artifact. However, since individual topologies are not
observable, it might be more accurate to study the convergence of the sum of the individual topologies and its order-by-order behavior. This is shown in the lower panels of Fig.~\ref{fig:3Ncomparison} for neutron matter for nonlocal regulators~\cite{Tews:2012fj}.

The remaining two leading $3N$ couplings are called $c_D$ and $c_E$ and determine the $3N$ contact interaction, as well as the one-pion-exchange--contact interaction. The LECs $c_D$ and $c_E$ are pure $3N$ couplings, meaning that they only appear in the $3N$ sector, and therefore have to be fit to systems with $A\geq3$. Up to now, Hamiltonians have been constructed using many different observables to constrain these couplings. 
Examples include fitting these couplings to $^4{\rm He}$ or $^3{\rm H}$ ground-state energies or charge radii, n-$\alpha$ scattering, $^3{\rm H}$ $\beta$ decay, or even to properties of many-body systems like $^{16}{\rm O}$, see Sec.~\ref{sec:fitting}. 
However, different determinations lead to very different results. For example, the ${\rm NNLO}_{\rm sat}$ interaction was fitted to properties of medium-mass nuclei and describes binding energies and charge radii sufficiently well, but produces a very small symmetry and neutron-matter energy. 
On the other hand, interactions fit to few-body systems produce good values for neutron matter and the symmetry energy, but seem to not describe nuclear charge radii or binding energies well.  
Hence, the uncertainty in these low-energy couplings leads to sizable uncertainties over the range of nuclear systems. Possible explanations for this varying behavior could be regulator artifacts (local vs. nonlocal regulators), a recently discovered derivation error for currents that have been used when fitting to $^3{\rm H}$ $\beta$ decay~\cite{Baroni:2018fdn}, or the omission of two-body operators for the charge radius.

We would like to emphasize here that it is imperative that a broader range of chiral EFT interactions is explored in nuclear-structure calculations and calculations of matter. 
Most importantly, these calculations need to provide meaningful estimates for theoretical uncertainties as well as an order-by-order analysis of the computed
observables, so that additional shortcomings can be identified and solutions may be found. 
There is currently a significant effort in the community to resolve these various limitations and it is the goal of this review to highlight different approaches and identify possible ways forward. 

\section{Constraining Nuclear Forces with Few- and Many-Body Observables}~\label{sec:observables}

\noindent
There exists plenty of data from decades of
experimental studies of the bound, resonant, and
continuum spectra in atomic nuclei throughout the nuclear chart. 
Likewise, extensive efforts have been put into the
construction of $NN$ and $3N$ forces to  higher orders in
chiral EFT~\cite{Reinert:2017usi,Entem:2017gor,Epelbaum:2014sza,Hebeler:2015wxa},
and with explicit inclusion of $\Delta$ isobars up to
N$^3$LO~\cite{PhysRevC.98.014003} for the longest-range parts. An important and
remaining challenge is how best to exploit the available experimental data to compare and
test the validity of proposed theories. This challenge encompasses the
problem of estimating the probability distributions of the numerical values of the LECs, especially in the presence of model discrepancies, i.e., EFT truncation errors. In this section, we will begin by discussing the present status of mathematical optimization of LECs in chiral EFT, and the consequences of different fitting strategies.

In the early days, the LECs, or general
parameters of the potential model, were mainly fitted to reproduce the world
database of $NN$ scattering cross sections up to $350$~MeV lab scattering energy.
Many-body forces were usually constrained to reproduce bulk observables in $A\geq3$ systems, once the LECs of the $NN$ force
were determined. This way, the fit of the LECs were largely driven by the 
$\sim$ 6000 $NN$ scattering observables in the scattering database. 

The corresponding objective
function $f(\mathbf{x})$, traditionally referred to as the $\chi^2$-function, against which nearly all
interaction models were calibrated and benchmarked can be
written as
\begin{equation}
  f(\mathbf{x}) = \sum_{g=1}^{N_g} \stackrel[\nu_{g}]{}{\rm min}\left\{ \sum_{i=1}^{N_{d,g}} \left( \frac{\nu_{g}\mathcal{O}^{\rm model}_{g,i}(\mathbf{x}) - \mathcal{O}^{\rm experiment}_{g,i}}{\sigma_{g,i}} \right)^2 + \left( \frac{1-\nu_{g}}{\sigma_{g,0}}\right)^2\right\}.
\label{eq:fobj}
\end{equation}
Here, the unknown vector of LECs in the interaction model is denoted by $\mathbf{x}$. The data consists of $N_g$ groups, often measured during the same experimental run, where the number of data points in each group $g$ is denoted $N_{d,g}$. The normalization constant $\nu_{g}$, together with its uncertainty $\sigma_{g,0}$, represents the systematic uncertainty of the measurements in group $g$. For an absolute measurement, the normalization is given by
$\nu_g=1 \pm 0$. Usually this means that the statistical and
systematic errors have been combined. Certain
experiments are not normalized at all. Instead, only the angular or
energy dependence of the cross section was determined. For such,
so-called floated data, $\nu_g$ is solved for by minimizing the
discrepancy between the model prediction $\mathcal{O}_{g,d}^{\rm
  model}$ and the experimental data points $\mathcal{O}_{g,d}^{\rm
  experiment}$. For practical purposes, the normalization error can be
considered infinite in such cases.

Most phenomenological interaction models, and early chiral EFT models were fitted to minimize the $\chi^2/\textrm{datum}$ in Eq.~\ref{eq:fobj} and only potentials with $\chi^2/\textrm{datum} \approx 1$ and certainly $\chi^2/\textrm{datum}<2$ were considered in nuclear-structure calculations. In general, pitfalls such as overfitting were seldom or never considered. These, and other problems pertaining to parameter estimation, were of minor concern until about a decade ago when \textit{ab initio} methods became capable of calculating heavier nuclei with controlled approximations. 

The canonical objective function in Eq.~\ref{eq:fobj} neglects all sources of theoretical uncertainty $\sigma_{\rm th}$, or model discrepancy in general, and any associated covariance structure. Nowadays much of the focus is shifted towards how interactions are optimized and how theoretical uncertainties are quantified. We will return to these points in detail later. For now, we focus on the experimental data that is typically included in the objective function.

Since EFTs are low-energy theories, not only $NN$ scattering observables, but also any experimental data from the low-energy spectrum of the nuclear Hamiltonian can be included in the
objective function. Also, since chiral EFT is an expansion in the small ratio $\{m_{\pi},p\}/\Lambda_b$, and not in the number of nucleons, $NN$ and many-nucleon interactions might be optimized simultaneously. 
The chiral effective Lagrangian also offers an advantageous link between $NN$ forces and low-energy $\pi N$ scattering. This makes it possible to use $\pi N$ scattering data to constrain the $\pi N$ LECs ($c_i$, $d_i$, and $e_i$), which enter the subleading $NN$ two-pion-exchange interaction, as well as the leading $3N$ two-pion exchange, and the currents. 
For many years, the uncertainties in the values of the $\pi N$ LECs were the largest source of parametric uncertainty. The measured $\pi N$ scattering data was neither abundant nor precise enough to sufficiently constrain all directions of the covariance matrix.

Serious effort has been invested into pinning down the $\pi N$ LECs. The relevant LECs were recently precisely determined using
the Roy-Steiner formalism~\cite{Hoferichter:2015tha, Hoferichter:2015hva}. Furthermore, a set of LEC values has been provided that captures the essential physics when working at different chiral orders. 
These values are reported in Table~I of Ref.~\cite{Siemens:2016jwj}, for different PC schemes in the heavy-baryon formalism and in the covariant formalism, and for
different chiral orders from NLO to N$^3$LO. Furthermore, results are
given in the $\Delta$-less and $\Delta$-full theories. Such a precise determination of the pion-nucleon LECs eliminates a sizable source of uncertainty, also in the $3N$ sector, see, e.g.,
Refs.~\cite{Hebeler:2009iv, Tews:2012fj} for specific examples in neutron matter.

When comparing the order-by-order series of the $\pi N$
scattering amplitudes in $\Delta$-less and $\Delta$-full chiral EFT in
the heavy-baryon formalism, the convergence seems to be improved when the
$\Delta$ degree of freedom is explicitly included, due to the large $\Delta$ loops at N$^3$LO in the $\Delta$-less theory. In $\Delta$-less chiral EFT, the $\pi N$ LECs absorb the effects of the $\Delta$ resonance in $\pi N$ scattering, leading to an enhancement of the LECs to unnaturally large values for, e.g., the $c_i$. The order-by-order
scattering amplitudes seem to prefer the $\Delta$-full theory with more natural LEC values. 
When using a covariant formulation, on the other hand, the convergence improves and the scattering amplitudes show excellent agreement with the Roy-Steiner analysis at N$^3$LO. However,
there is no rigorous PC argument for a covariant
formulation. 

Whether data generally support the
$\Delta$-full theory over the $\Delta$-less theory needs to be studied in side-by-side
comparisons. In this context, it is also important to 
identify the relevant momentum scales in processes that include the $\Delta$, i.e., the $\Delta$-nucleon mass splitting, $\Delta m=m_{\Delta}-m_N\approx 300$ MeV or the $\Delta$ production
scale,$\sqrt{m_N(m_N-m_{\Delta})}\approx 500$ MeV.

Having specified the pion-nucleon LECs, one needs to constrain the
remaining short-range LECs in the $NN$ interaction. The fitting
of $NN$ interactions to phase shifts simplifies the fitting procedure,
since certain parts of the $NN$ contact interaction only act in specific partial waves. However, phase shifts are not experimental data but model-dependent quasi-data. 
Therefore, fitting directly to scattering data is preferred, which also allows for experimental uncertainties to be directly used in the likelihood function.
Also, when fitting directly to data, all remaining  model dependence stems from the nuclear Hamiltonian. In addition, it has been shown that interactions that reproduce the phase shifts do not necessarily have to reproduce
scattering data as well~\cite{Stoks:1993zz} and a simple reproduction
of phase shifts is not sufficient to reproduce experimental data from few- and many-body sectors. 
This has also been observed in nuclear lattice calculations~\cite{Elhatisari:2016owd}.

\begin{figure}[t]
\centering
\includegraphics[trim= 5.0cm 1cm 6cm 6cm, clip=,width=0.65\columnwidth]{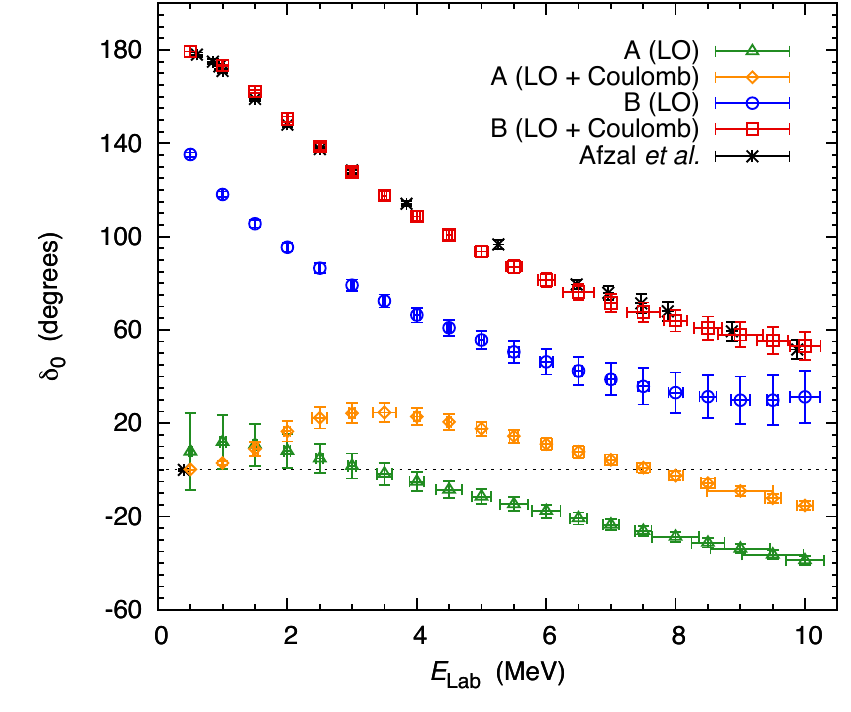}
\caption{\label{fig:AlphaPhaseShiftLattice} The $\alpha-\alpha$ scattering
  phase shifts for two nuclear lattice interactions A and B, which have
  nearly identical $NN$ phase shifts and three- and four-nucleon bound
  states. However, these interactions show different behavior in
  multiple-$\alpha$ nuclei: Interaction A leads to a Bose gas of
  $\alpha$ particles, while interaction B provides bound
  nuclei. Instead of testing these interactions directly in many-body
  calculations, additional information on $\alpha$-$\alpha$ scattering
  phase shifts can help distinguish between the two interactions. In
  this case, interaction B correctly describes the experimental
  data, while interaction A fails.  Reprinted figure with permission from Ref.~\cite{Elhatisari:2016owd}. Copyright 2016 by the American Physical Society.}
\end{figure} 

There are several pieces of data that might improve the quality of 
nuclear Hamiltonians. For example, several studies have found that nuclear interactions are very
sensitive to scattering observables such as $n-\alpha$, 
$\alpha-\alpha$, or $N-d$ scattering~\cite{Lynn:2015jua, Elhatisari:2016owd}. For example, local chiral $NN$ and $3N$ forces, 
when constrained using the \isotope[4]{He} ground-state 
energy and $n-\alpha$ scattering, show an excellent reproduction 
of nuclei up to \isotope[16]{O}~\cite{Lonardoni:2017hgs, Lonardoni:2018nob} while previous local interactions, which were fit to other observables, failed to reproduce charge radii\footnote{One must, however, note that LECs $c_E$ and $c_D$ of these previous local interactions were 
fit to the triton binding energy and $\beta$-decay half 
life~\cite{Gazit:2008ma}, which suffered from a missing factor $-1/4$ when matching $c_D$ in the current with $c_D$ in the $3N$ force. This 
could explain the bad description of nuclear charge radii.}. 

Another excellent illustration of the benefit of including information 
from scattering observables is provided by a calculation of Ref.~\cite{Elhatisari:2016owd} in the 
framework of nuclear lattice EFT~\cite{Lee:2008fa} using two 
Hamiltonians that provide nearly identical $NN$ phase shifts and 
binding energies of $A=3,4$ nuclei. However, for multiple-$\alpha$
nuclei, it was found that one interaction leads to a Bose condensate of
$\alpha$ particles while the other interaction leads to more realistic
binding energies for $A>4$ nuclei. Additional information on 
$\alpha-\alpha$ scattering phase shifts helps distinguish 
between the two interactions, see Fig.~\ref{fig:AlphaPhaseShiftLattice}: only one of the two interactions correctly describes experimental 
data. 

\begin{figure}[t]
\centering
\includegraphics[trim= 0 0 0 0, clip=,width=0.6\columnwidth]{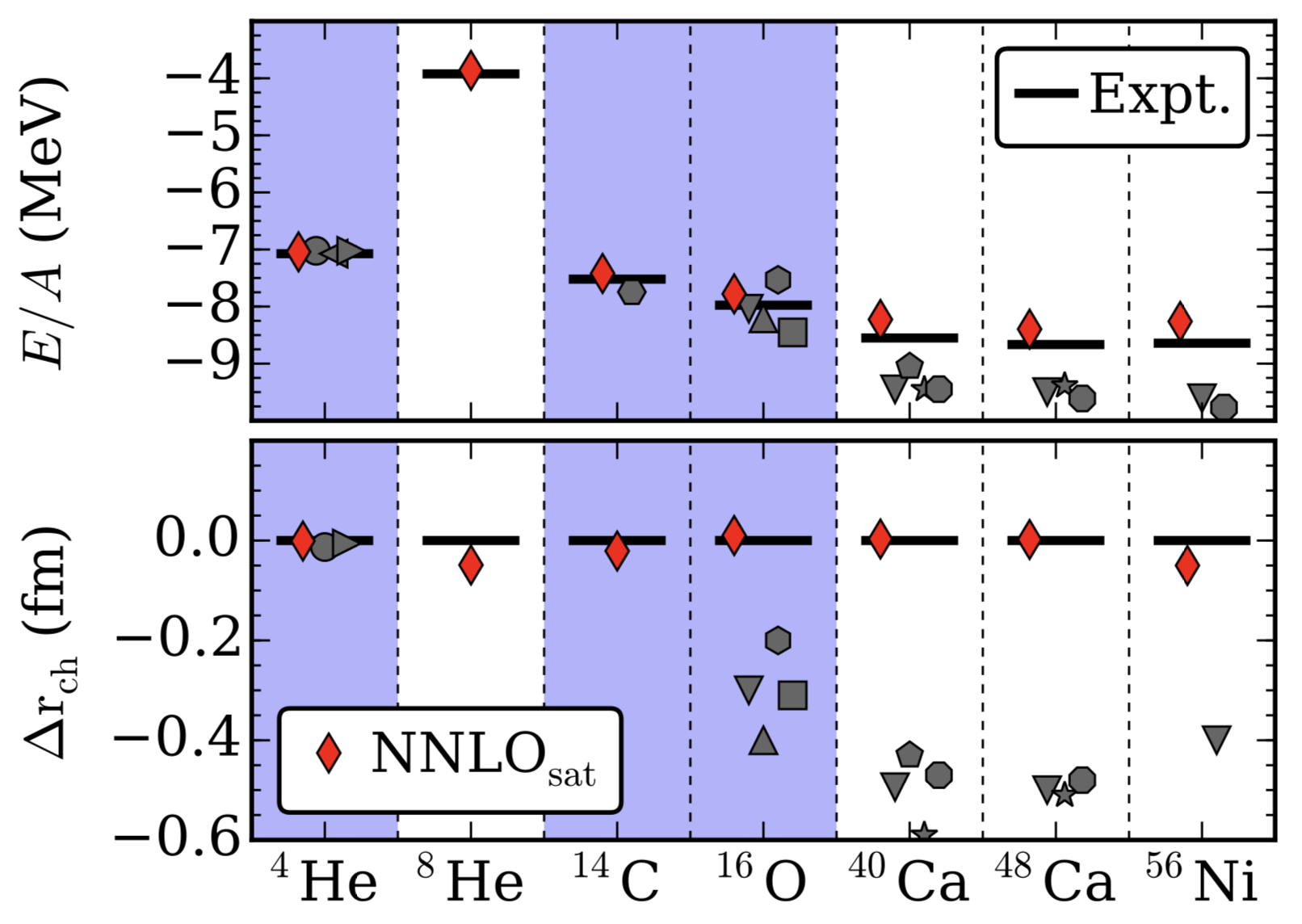}\\
\caption{\label{fig:nnlo_sat} Ground-state energies per nucleon (top) and differences between theoretical and experimental charge radii (bottom) for selected light and medium-mass nuclei and results from \textit{ab initio} computations. The red diamonds mark results based on the chiral interaction NNLO$_{\rm sat}$. The blue columns indicate which nuclei were included in the optimization of the LECs in NNLO$_{\rm sat}$, while the white columns are predictions. Grey symbols indicate other chiral interactions.
Reprinted figure with permission from Ref.~\cite{Hagen_2016}.}
\end{figure}

In addition to scattering observables, it is possible
to include many-body observables in the objective function, i.e., to 
fit both $NN$ and $3N$ interactions
simultaneously to $NN$ scattering cross sections, few-nucleon data, 
and binding energies and radii of selected isotopes of, e.g., carbon 
and oxygen. 
This approach has been pursued in the construction of the 
NNLO$_{\rm sat}$ potential~\cite{Ekstrom:2015rta} with the goal of
developing predictive \textit{ab initio} capability for light and 
medium-mass nuclei, see Fig.~\ref{fig:nnlo_sat}. This optimization 
strategy can be extended to include, e.g., saturation properties of 
nuclear matter or additional few- or many-body scattering observables. 
The advantages of such a procedure is that higher-density behavior 
of nuclear interactions, most importantly nuclear saturation, as well 
as two- and three-nucleon interactions are included in the fit.
Resulting predictions with NNLO$_{\rm sat}$ have generally produced accurate 
charge radii and good, but somewhat underbound, ground-state energies. Spectra, however, are often not well reproduced~\cite{Liu:2018mtu,Randhawa:2019dxq}, pointing to the need for further refinement. 

The question of whether or not to exploit medium-mass nuclei for fitting might
invite different viewpoints. Recently, however, the many-body
community has shown a shift in thought to a new mindset, away from a
more reductionist desire where two-body interactions should be
informed solely from two-body experimental inputs and three-body
interactions should be informed solely on three-body experimental inputs. It seems that consensus has been reached that using
many-body observables in fits of nuclear Hamiltonians is a useful
approach given current uncertainties in the interactions, although it still remains unclear how to avoid capturing errors from the particular many-body method in such fits. For now, this requires specific domain knowledge about the choice of the many-body method for critical observables.

\begin{figure}[t]
\centering
\includegraphics[trim= 0 0 0 0, clip=,width=0.45\columnwidth]{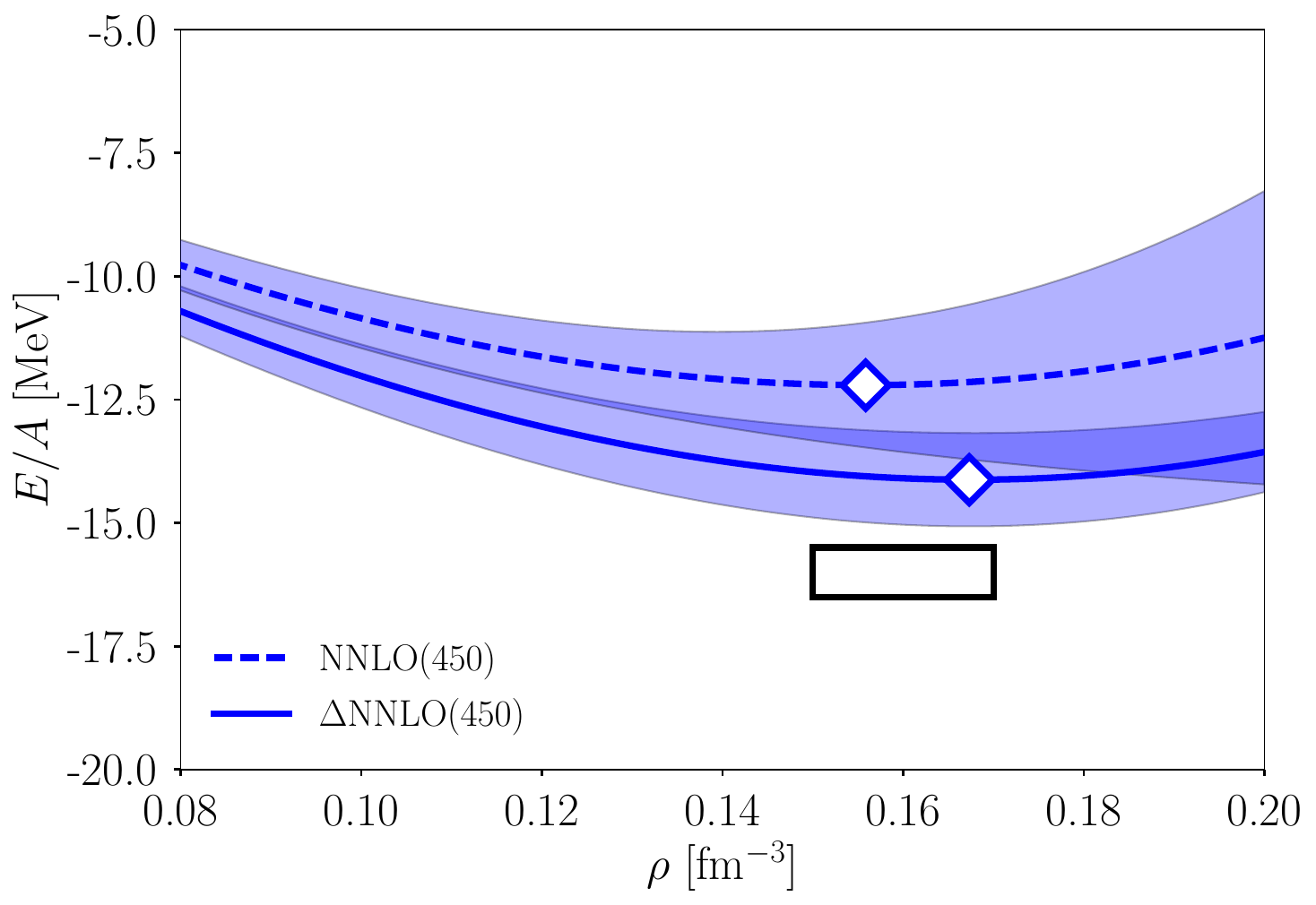}
\includegraphics[trim= 0 0 0 0, clip=,width=0.52\columnwidth]{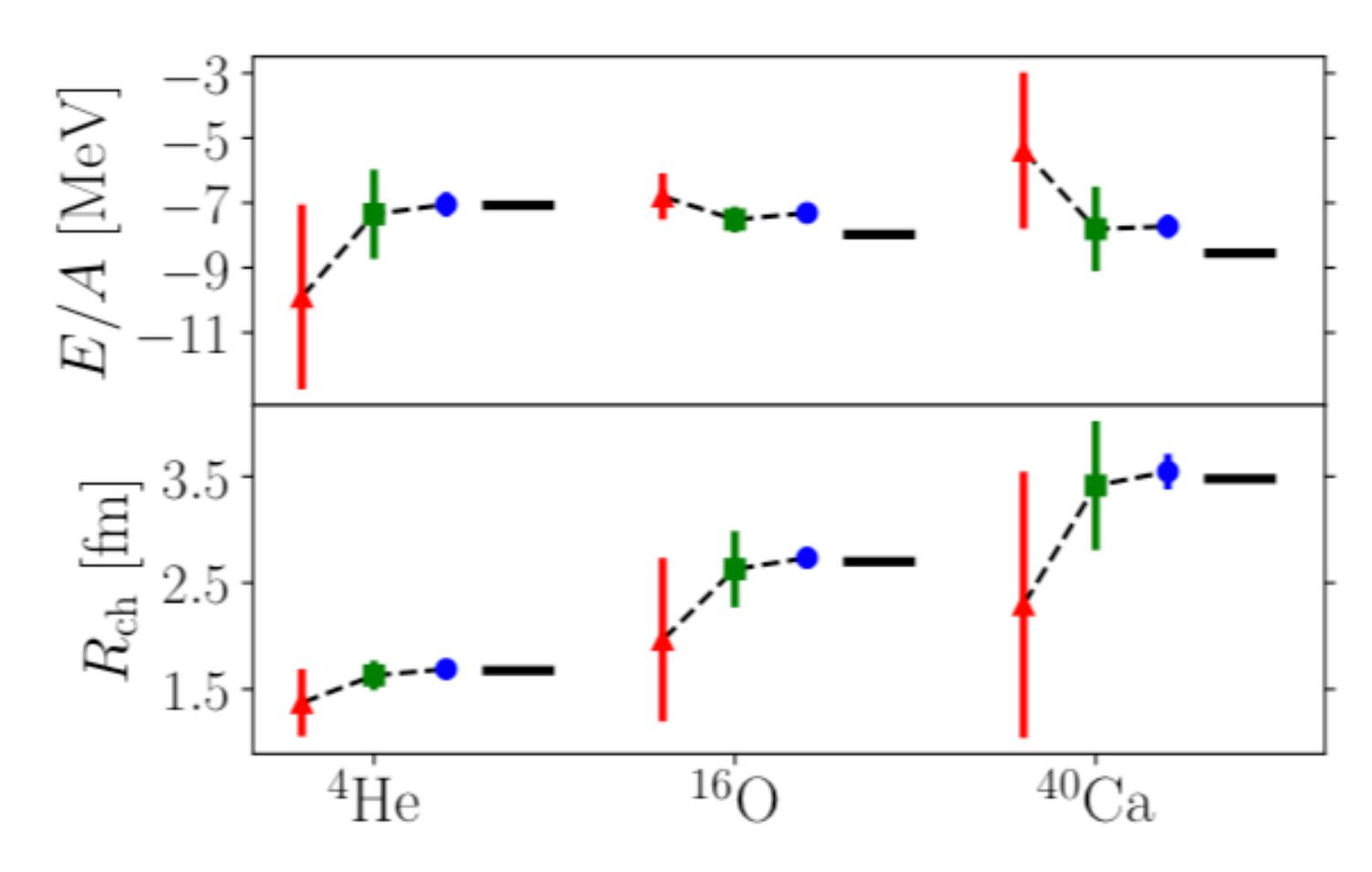}
\caption{\label{fig:deltafull} [Left panel] Coupled-cluster calculations of the energy per nucleon (in MeV) in symmetric nuclear matter at N$^2$LO in chiral EFT with (solid line) and without (dashed line) the $\Delta$
isobar. Both interactions employ a momentum-regulator cutoff
$\Lambda$ = 450 MeV and were fitted to the same few-nucleon data only. 
The shaded areas indicate the estimated EFT truncation errors following 
the prescription presented in Ref.~\cite{Furnstahl:2015rha}. The 
diamonds mark the saturation point and the black rectangle indicates 
the region $E/A$ = −16$\pm$ 0.5 MeV and $\rho$ = 0.16 $\pm$ 0.01 
fm$^{−3}$. 
[Right panel] Ground-state energy per nucleon and charge radii for selected 
nuclei computed with coupled-cluster theory and the $\Delta$-full 
potential from Ref.~\cite{Ekstrom:2017koy}. For each nucleus, the points from
left to right correspond to LO (red triangle), NLO (green square), and 
N$^2$LO (blue circle). 
Reprinted figures with permission from Ref.~\cite{Ekstrom:2017koy}. Copyright 2018 by the American Physical Society.
}
\end{figure}

From a practitioner's viewpoint, the inclusion of many-body information 
in the fit of nuclear forces seems to be important to properly 
describe nuclear-structure observables, like radii and binding energies, and is very promising for the prediction of driplines etc.
In contrast, it remains to be seen whether it is possible to formulate a nuclear-interaction model with just a few
parameters that can be fit solely to corresponding few-nucleon data. 
It should be pointed out, though, that the $\Delta$-full NLO and N$^2$LO 
interactions presented in Ref.~\cite{Ekstrom:2017koy} provide a relatively good 
description of binding energies and charge radii of several 
medium-mass nuclei and decent saturation properties of nuclear 
matter, see Fig.~\ref{fig:deltafull}. In this context, it is noteworthy that all 
LECs were fitted only to reproduce binding energies and radii of 
nuclei with $A \leq 4$ , $NN$ scattering phase shifts up to 200 MeV, 
and using $\pi N$ LECs inferred from the Roy-Steiner analysis. 
This result might emphasize the importance of the $\Delta$ degree of freedom in chiral EFT. 

\begin{figure}[t]
\centering
\includegraphics[trim= 0.5cm 1.5cm 12.5cm 5cm, clip=,width=0.695\columnwidth]{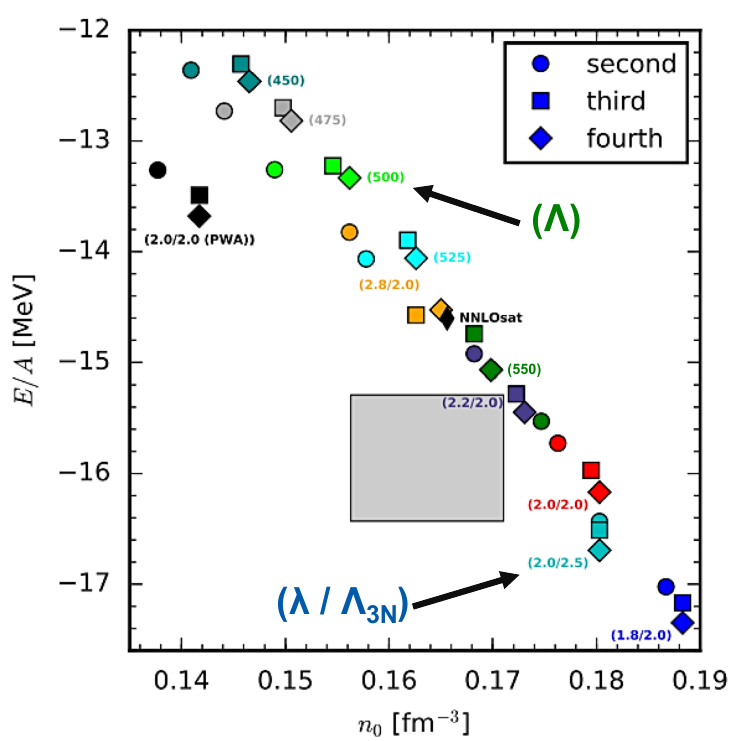}
\caption{\label{fig:MBPTsat} Saturation point for several chiral
  Hamiltonians of
  Refs.~\cite{Hebeler:2010xb,Ekstrom:2015rta,Carlsson:2015vda} in
  nuclear matter at second, third, and fourth order in many-body
  perturbation theory. For comparison, the saturation point of
  NNLO$_{\rm sat}$ is also shown. The interactions fall on a
  Coester line that barely touches the empirical saturation
  point (gray region). 
  Reprinted figure with permission from Ref.~\cite{Drischler:2017wtt}. Copyright 2019 by the American Physical Society.}
\end{figure}

In general, when studying the saturation properties of symmetric 
nuclear matter, conventional chiral interactions, which do not 
reproduce properties of medium-mass nuclei very well, also fail 
to describe the empirical saturation point, both in density and
binding energy. Generally speaking, the saturation density of a 
given Hamiltonian affects nuclear radii while the saturation 
energy affects the binding energies. While this argument is heuristically based, the saturation properties are generally a good 
indicator for the ability of an interaction to describe masses and radii of nuclei and vice versa. Indeed, the NNLO$_{\rm sat}$ 
interaction~\cite{Ekstrom:2015rta} is fit to radii and binding energies of selected medium-mass nuclei up to $A=24$ and saturates at the correct density, see Fig.~\ref{fig:MBPTsat}. 
\begin{figure}[t]
\centering
\includegraphics[trim= 0 0 0 0, clip=,width=0.4\columnwidth]{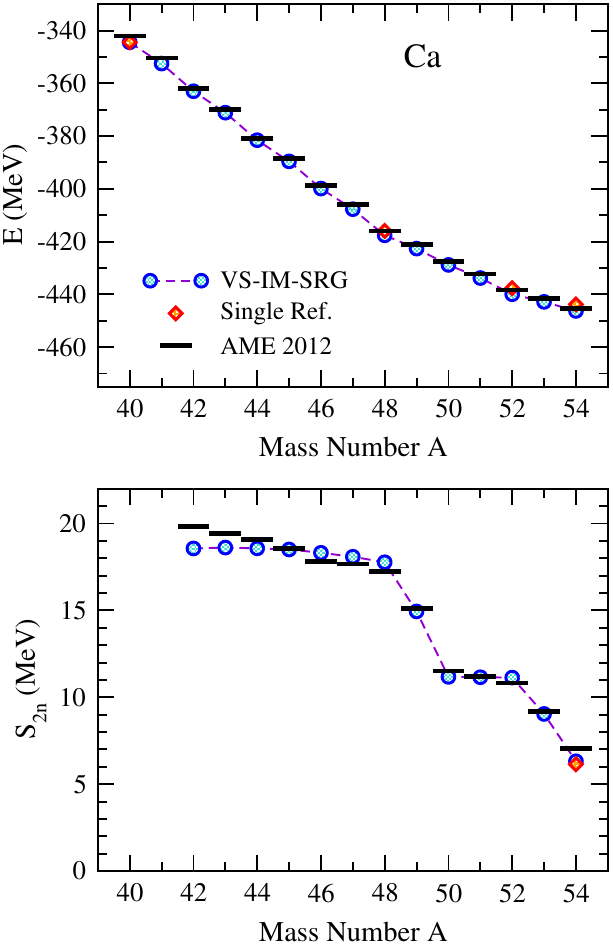}
\includegraphics[trim= 0 0 0 0, clip=,width=0.408\columnwidth]{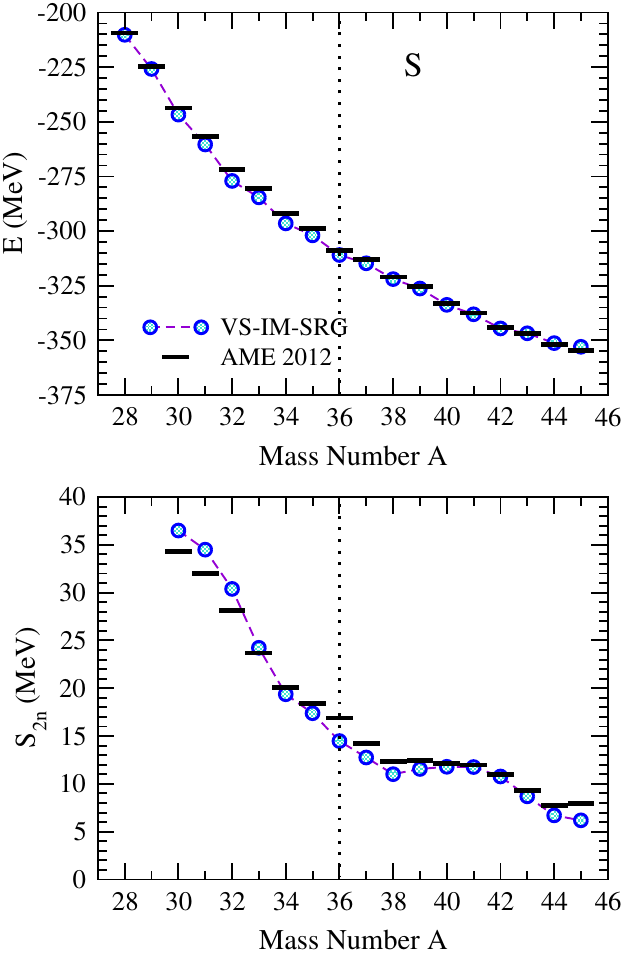}\\
\includegraphics[trim= 0 0 0 0, clip=,width=0.4\columnwidth]{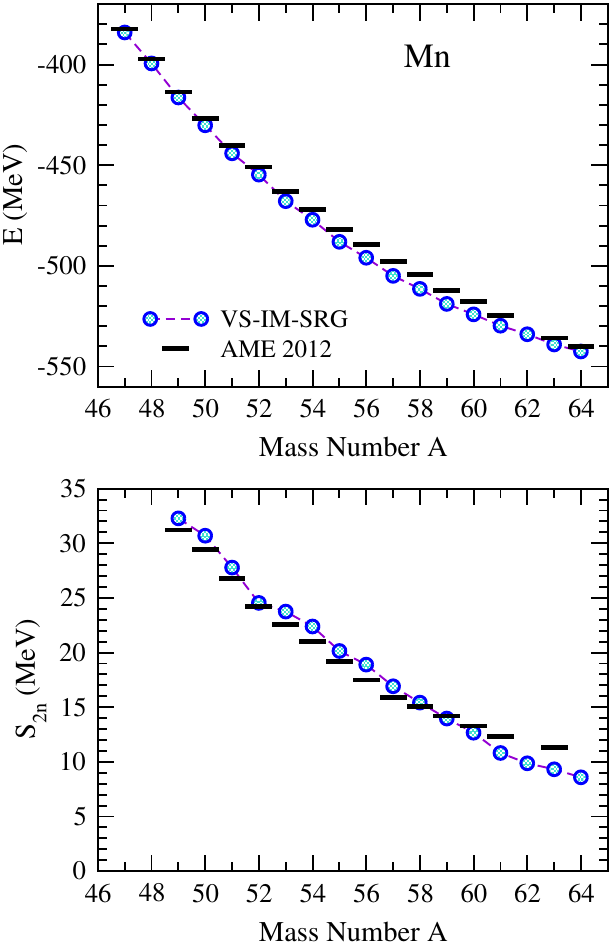}
\includegraphics[trim= 0 0 0 0, clip=,width=0.4\columnwidth]{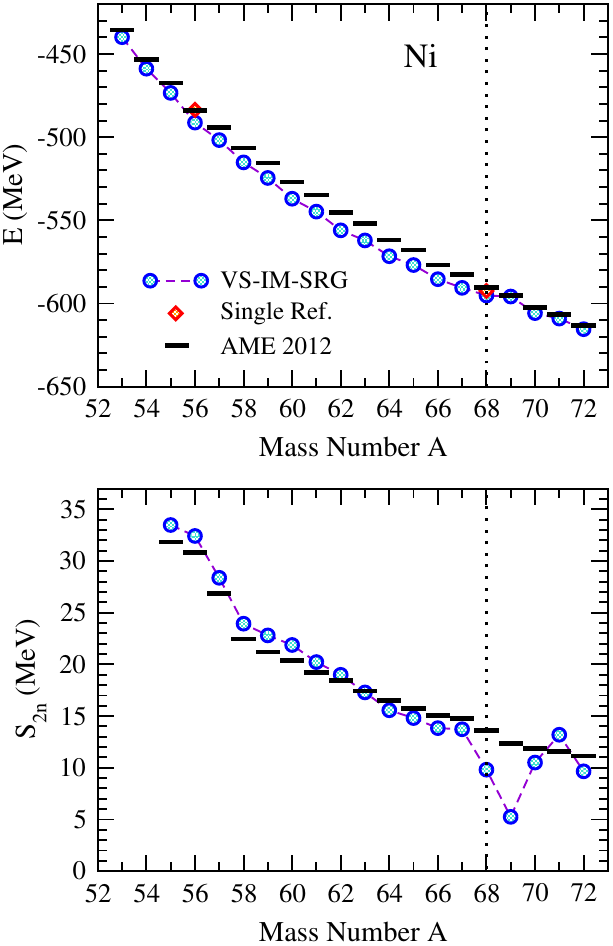}
\caption{\label{fig:EM1.8-2.0}
Ground-state energies and two-neutron separation energies for Ca, S, Mn, and Ni as functions of the mass number for the EM 1.8/2.0 interaction of Ref.~\cite{Hebeler:2010xb}. 
Reprinted figures with permission from Ref.~\cite{Simonis:2017dny}. Copyright 2017 by the American Physical Society.}
\end{figure} 

Nevertheless, the energy-per-particle ($E/A$) of the NNLO$_{\rm sat}$
interaction does not fully coincide with the empirical saturation
point. The same is true for the EM 1.8/2.0 Hamiltonian of
Refs.~\cite{Hebeler:2010xb,Simonis:2015vja}, that saturates at too high densities and
energies but describes nuclear energies extremely well through the tin region~\cite{Morris:2017vxi}. Figure~\ref{fig:EM1.8-2.0} shows the ground-state energies and two-neutron separation energies for this Hamiltonian in four isotopic chains, exhibiting an excellent reproduction of the experimental data. Figure~\ref{fig:MBPTsat} displays the saturation properties of these and
additional chiral Hamiltonians of Refs.~\cite{Ekstrom:2015rta,
  Carlsson:2015vda,Hebeler:2010xb}. These Hamiltonians form a
Coester-like line, that barely touches the empirical saturation
point. Hence, the question arises if the empirical saturation point,
determined by averaging over a large set of density functional
theories (DFTs) that have been adjusted to energies and radii of
nuclei up to lead, is model dependent, i.e, if this ``box'' remains the
same for \textit{ab initio} calculations and DFTs, and what the corresponding uncertainties are. For instance, surface terms in \textit{ab initio}
calculations can have a different magnitude  compared to
DFTs~\cite{Buraczynski:2016jia}, which might influence the bulk terms and therefore the saturation point.

Another interesting observation is that nuclear interactions that
have an excellent reproduction of properties of light to
medium-mass nuclei and the saturation point, seem to produce insufficient repulsion in pure neutron matter, leading to a too small symmetry
energy and a too soft equation of state (EOS) for neutron-rich matter. This is true for both phenomenological interactions, like for the AV18+IL7
Hamiltonian, as well as for some chiral Hamiltonians like
NNLO$_{\rm sat}$. On the other hand, most interactions that
describe neutron matter well, seem to be less accurate in other many-body systems. An exception is given by local chiral interactions that have been 
fit also to $n-\alpha$ scattering and nicely reproduce both 
properties of nuclei up to \isotope[16]{O} and neutron matter~\cite{Lynn:2015jua, Lonardoni:2017hgs}.

Several additional constraints might be used to determine nuclear
interactions. First, one could match nuclear Hamiltonians to lattice-QCD calculations. For instance, when putting neutrons in a spatial box and
squeezing the box, comparisons of many-body calculations of this
system with lattice-QCD calculations may be possible at physical
pion masses. We will discuss these possibilities in
Sec.~\ref{sec:lqcd} and, instead, focus here on nuclear lattice
calculations~\cite{Lee:2008fa}. 
It might be possible to fit continuum nuclear Hamiltonians to nuclear lattice data if the lattice spacing can be reduced. Work along these lines is in progress. 
Furthermore, it might be interesting to use finite-temperature data from nuclear lattice simulations to constrain chiral forces in finite-temperature systems, because for some many-body methods, finite-temperature observables are easier to compute. 
For which systems such an approach might be useful is still an open question.

\begin{figure}[t]
\centering
\includegraphics[trim= 0 0 0 0, clip=,width=0.6\columnwidth]{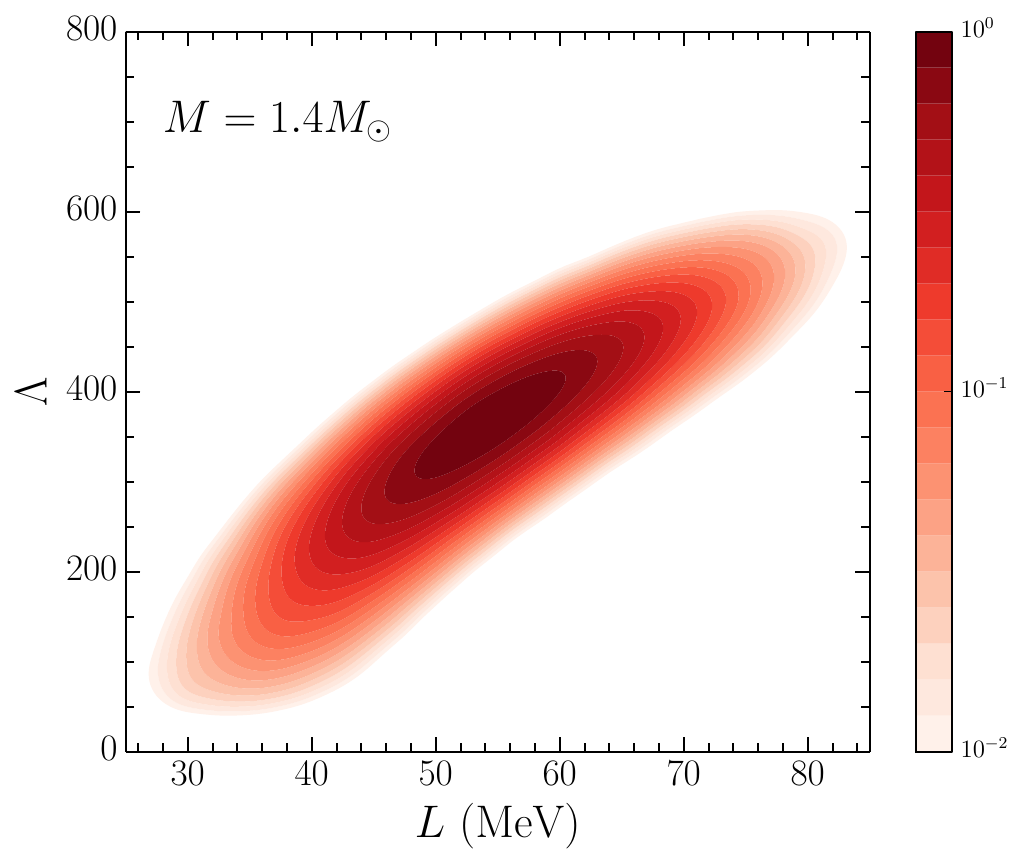}
\caption{\label{fig:Lambda-L}
Correlation plot for the tidal polarizability of a $1.4 M_{\odot}$ star and the slope parameter of the symmetry energy $L$. 
Reprinted figure with permission from Ref.~\cite{Lim:2018bkq}. Copyright 2018 by the American Physical Society.}
\end{figure} 

Finally, additional constraints might come from nuclear
astrophysics, and in particular neutron stars. Neutron stars explore 
nuclear matter at densities around saturation density and above, 
and might be helpful to constrain nuclear interactions that enter
neutron-star physics via the EOS. 
The EOS is currently highly uncertain, but fascinating observations of neutron-star mergers~\cite{Abbott:2017}, as well as anticipated 
precise neutron-star radius measurements, will shed more light on 
the EOS of neutron-star matter~\cite{Annala:2017llu, Most:2018hfd, Lim:2018bkq,Tews:2018iwm}. Constraints on the EOS of neutron stars might then be used to constrain the nuclear interaction itself. 

In Ref.~\cite{Lim:2018bkq} it was shown that, if no phase transition 
occurs within typical neutron stars with masses around 
$M=1.4 M_{\odot}$, then the so-called empirical $L$ parameter, which is defined as the slope of the symmetry energy at saturation density and, thus, is related to the pressure of neutron matter, can be 
constrained by observing the tidal polarizability of a neutron star 
within a neutron-star merger. Figure~\ref{fig:Lambda-L} displays this correlation. The tidal polarizability $\lambda$ describes how a neutron star deforms under an external gravitational 
field created by a second neutron star in a binary, 
$Q_{ij}=\lambda E_{ij}$, where $Q_{ij}$ is the quadrupole moment and
$E_{ij}$ is the external gravitational field. Typically, a 
dimensionless tidal polarizability is defined, $\Lambda=\lambda/M^5$. 
The tidal parameter appears as a post-Newtonian fifth-order correction 
to the wave-front phase~\cite{Flanagan:2007ix, Damour:2009vw} and is, thus,
difficult to constrain. A precise measurement, however, may help to
constrain the symmetry energy slope parameter $L$ and, thus, help to
pin down nuclear interactions around saturation density. In addition 
to neutron-star mergers, precise radius measurements are expected to become available in the near future. 
Therefore, it is likely 
that the nuclear EOS of neutron-star matter in the density
range of $1-2 n_{\rm sat}$ could be pinned down within the next few years. This density range can be explored by current \textit{ab
  initio} calculations and, therefore, certain Hamiltonians might be
excluded based on their neutron-star-matter predictions.

At this point, we stress that a theoretical model can generally be
optimized to any set of data if the theoretical uncertainty is
properly accounted for. In this sense, all data is equally good as
long as uncertainties are meaningful. However, the energies to which a Hamiltonian has to be fit should neither be too low nor too high and should be chosen in a regime where chiral EFT works reasonably well. 
Nevertheless, it is a theoretical and computational challenge 
to incorporate expected theory errors \textit{a priori}. One reason is that
it is not entirely clear how to determine the relevant momentum 
scale of an observable, e.g., a nuclear bound state. More work is needed in this direction.

\section{Improving Two- and Many-Body Nuclear Forces with Novel Fitting Strategies and Including Higher Orders in Chiral EFT}\label{sec:fitting}

In this section, we will focus on nascent approaches to the problem 
of inferring the probability distributions of the LECs in EFT 
descriptions of the nuclear force. Parameter estimation is a problem 
of a very general nature that appears in most branches of science. To 
calibrate the parameters of a theory typically requires inductive 
reasoning in the presence of incomplete information. In practice, 
parameter estimation requires a mixture of probability theory, 
applied mathematics, statistical analysis, and most importantly, 
expert knowledge about the underlying theory and the pool of 
calibration data. A framework for parameter estimation and model 
comparison is provided by Bayesian statistics via 
Bayes' theorem. It informs one how to express 
probabilities for a hypothesis $H$, e.g., model parameters or a model itself, given some dataset $D$
\begin{equation}
  P(H|D) = \frac{P(D|H)P(H)}{P(D)}\,.
\end{equation}

To utilize one of the main advantages of EFTs, all theoretical
predictions should be carried out order-by-order. This exposes a crude
estimate of the uncertainty coming from the EFT truncation. The
possible systematic uncertainties of the $k/\Lambda_b$ expansion is one example of the
so-called prior knowledge (or belief), i.e. something we know or 
have reasons to believe to be true even before we have looked at 
new data. This is valuable information that should be retained in, 
e.g., an estimation of the possible values of the LECs. Another 
possible prior could be based on the expectation that the LECs 
should have natural values when expressed in appropriate powers 
of the breakdown scale.

The application of Bayesian statistics in \textit{ab initio} nuclear 
theory is a recent and important advancement, see e.g., 
Refs.~\cite{Melendez:2017phj,Wesolowski:2018lzj}. A significant step towards Bayesian 
parameter estimation in chiral EFT was taken in 
Ref.~\cite{Melendez:2019izc} where the EFT truncation error was modeled with a Gaussian process. This process can be incorporated in
a likelihood for data (observables) 
correlated across an independent variable, e.g., the scattering angle in a differential cross section, as part of future parameter estimation calculations. 

In some cases, a Bayesian analysis would yield results similar to a frequentist analysis, for instance in the presence of 
overwhelmingly abundant data. Nevertheless, a Bayesian analysis presents several advantages. It enables 
an extraction of the probability distribution for the parameters of 
a model. It also makes all prior assumptions very clear. It further allows for a straightforward way for formulating hypothesis testing, and contains explicit mechanisms for guarding against overfitting. 
Although Bayesian inference is straightforward technically, the 
Bayesian approach comes at the expense of heavy computations via involved Markov chain Monte Carlo evaluations (MCMC) of posterior 
probability distributions. Although this can be alleviated to a 
certain extent using modern high-performance computing resources, 
the exponential cost induced by the curse of dimensionality quickly 
catches up, especially if the data types in the likelihood require 
expensive many-body calculations. Furthermore, MCMC itself is a complicated 
tool with challenging convergence diagnostics. Thus, future work should focus on (i) novel methods for constructing emulators or 
surrogate models to replace original expensive computation, e.g. 
Gaussian processes or methods from machine learning such as Bayesian 
optimization~\cite{Ekstrom_2019} or eigenvector continuation~\cite{Frame2018,Konig:2019adq,Ekstrom:2019lss}, (ii) efficient MCMC methods that 
scale well with an increasing dimensionality of the parameter domain, 
such as, e.g., Hamiltonian Monte Carlo, and (iii) methods for finding 
interesting subsets of the parameter domain in case of possible 
multi-modal and multi-dimensional posterior distributions, e.g., 
Approximate Bayesian Computation or History Matching~\cite{vernon2014}. 

As mentioned in Sec.~\ref{sec:observables}, since the early 2000s, 
state-of-the-art chiral $NN$ interactions have been constructed at 
N$^3$LO~\cite{Entem:2003ft,EGMN3LO} and fitted to minimize the 
$\chi^2/\textrm{datum}$ with respect to the world scattering $NN$ 
database. The importance of combining improved fitting algorithms, 
statistical inference, and expert knowledge of the theory is nowadays 
well recognized. For example, when first- and second-order 
derivative-based optimizing algorithms were applied to chiral 
interactions at N$^2$LO~\cite{Carlsson:2015vda}, four local minima 
were found when optimizing LECs in the $NN$ sector separately from 
the $3N$ sector. The canonical objective function consisted of the 
proverbial $\chi^2/\textrm{datum}$, augmented with a nascent estimate 
of the chiral EFT truncation error in the uncorrelated 
limit~\cite{Carlsson:2015vda,Wesolowski:2018lzj}. When simultaneously 
fitting $NN$ and $3N$ interactions, only one local minimum was found.

There exist other strategies for discriminating between local minima. 
For example, some minima lead to unnaturally large LECs, break Wigner 
symmetry~\cite{Mehen:1999qs}, or do not properly reproduce phase 
shifts in certain partial waves. In a Bayesian framework, such 
expectations could be imposed as priors already at the outset.
This reflects the importance of exploiting and soliciting expert 
knowledge about the theory. As another example, recent studies of 
N$^3$LO contact interactions showed a redundancy of LECs which can
be seen in terms of unitary transformations~\cite{Reinert:2017usi}
or in terms of statistical analyses~\cite{Wesolowski:2018lzj}. 
Properly accounting for this redundancy leads to softer potentials
with smaller Weinberg eigenvalues, allows for the elimination of 3 of the 15 
contact operators at N$^3$LO, removes multiple fit minima, and 
drastically improves the fits. 

While the situation with multiple fit minima at N$^2$LO and 
N$^3$LO can be understood and resolved, at higher orders similar situations 
may appear and several local minima may be found. This stresses 
the importance of identifying interesting subsets of the parameter 
domain and extracting the full posterior probability of the LECs 
as opposed to merely locating the maximum posterior location. The 
latter could very well hide the fact that most of the probability 
resides away from the posterior maximum. 

With these novel fitting strategies, it is then possible to access
higher orders in the chiral PC. We now discuss this 
second possible pathway. As was stated before, the main 
uncertainty in calculations of nuclear systems stems from the
truncation of the chiral expansion. Naturally, when assuming that 
all expansion coefficients in Eq.~(\ref{eq:EFTexpansion}) are of 
natural order, the uncertainty is reduced when performing calculations 
at higher orders in the chiral expansion, because $Q<1$. Recently, 
several $NN$ interactions at N$^4$LO or with additional N$^5$LO 
contact terms have been developed~\cite{Reinert:2017usi,Entem:2017gor}, which lead to an excellent reproduction of $NN$ scattering data. On 
the other hand, while there is tremendous progress in pushing the 
$NN$ interactions to higher orders, many-body forces are usually 
limited to N$^2$LO. As a consequence, $NN$ and many-body forces can not be treated consistently. 
While there have been calculations of nucleonic matter 
including both $NN$ and $3N$ interactions at 
N$^3$LO~\cite{Tews:2012fj, Drischler:2016djf,Drischler:2017wtt}, 
N$^3$LO $3N$ interactions have only recently been made available 
in the form of matrix elements to be used in calculations of finite 
nuclear systems~\cite{Hebeler:2015wxa} due to their complicated nature.
In this context, it would be interesting to answer the question of how
one can construct $NN$ interactions in such a way that complicated $AN$ forces are minimized. 
Can off-shell effects be traded for minimizing certain $3N$ topologies, e.g., the complicated N$^3$LO ring topologies, to push chiral interactions to higher orders in a more
feasible way?

\begin{figure}[t]
\centering
\includegraphics[trim= 0cm 1.5cm 2.0cm 2cm, clip=,width=0.65\columnwidth]{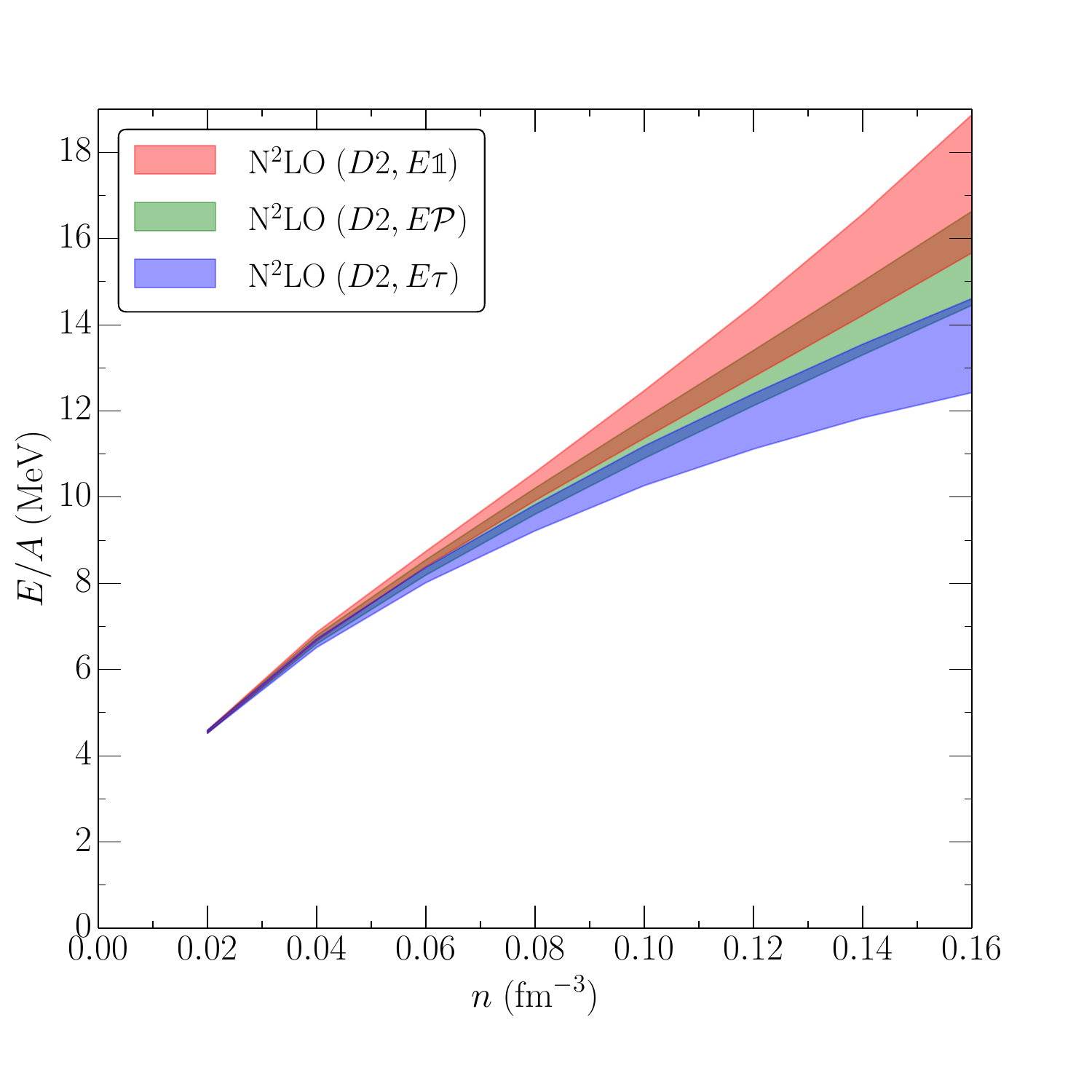}
\caption{\label{fig:ekmeosnm} Neutron matter energy per particle as a
  function of density for three different N$^2$LO $3N$ contact-operator choices for local regulators. Due to antisymmetry
  arguments, these different operator choices should be degenerate,
  but local regulators break this Fierz rearrangement freedom.
  Reprinted figure with permission from Ref.~\cite{Lynn:2015jua}. Copyright 2016 by the American Physical Society.}
\end{figure} 

Another crucial question is to estimate the challenges that are
presented by the regulator choice. Several regulators have been
employed in recent years: nonlocal regulators, as in 
Refs.~\cite{EGMN3LO,Entem:2003ft} for $NN$ or Ref.~\cite{vanKolck1994}
for $3N$ interactions, or local regulators, as in 
Refs.~\cite{Navratil:2007zn,Gezerlis:2013ipa,Piarulli:2014bda,Tews:2015ufa, Lynn:2015jua}.
Generally, a good choice is given by a regulator that 
minimizes the finite-cutoff artifacts in calculations. Such 
regulators should not affect the long-range behavior of nuclear 
interactions, which is a prediction of chiral EFT. Due to the 
locality of the long-range pion exchanges, a local coordinate-space 
regulator presents a good choice~\cite{Reinert:2017usi}. 
This has led to the development of semilocal regularization schemes, as in 
Ref.~\cite{Epelbaum:2014sza, Reinert:2017usi}. In that case, the 
short-range part of the potential will absorb the regulator artifacts 
and strongly depend on the regulator. 
A problem is that when naively applying the same long-range regulators in the $3N$ sector, the 
two-pion--one-pion-exchange N$^3$LO 3N interactions violates chiral symmetry and requires a counter term that breaks chiral symmetry. A solution is to use ``higher-derivative regularization'', see e.g. Ref.~\cite{Long:2016vnq}.

However, for these regularization schemes, additional problems have been observed. Chiral interactions do not lead to renormalization group (RG) invariance of observables when 
using Weinberg PC, that is, when varying the cutoff in 
a sufficiently large range, results do not remain cutoff
independent. Instead, spurious bound states appear in partial waves
with attractive tensor interactions~\cite{Nogga:2005hy}. For local
regulators, it has been observed that typical cutoff values lead to
lower effective cutoffs~\cite{Tews:2015ufa,Dyhdalo:2016ygz}, or that
local regulators violate the Fierz rearrangement freedom, that is,
results depend on the choice of the operator structure although
antisymmetricity considerations show that they should be
independent~\cite{Lynn:2015jua, Huth:2017wzw}. This is especially
relevant in the $3N$ sector, as shown in Fig.~\ref{fig:ekmeosnm}.

While these regulator artifacts are inversely proportional to the
cutoff, and, thus, vanish in the limit of high cutoffs, in the limited 
cutoff range employed for chiral EFT interactions, they usually play 
a role. The regulator artifacts are corrected when short-range
interactions at higher orders in the PC are included, as
demonstrated in Ref.~\cite{Huth:2017wzw}. Thus, pushing interactions
to higher orders might cure the most sizable regulator artifacts. In 
the $NN$ sector, when going to N$^3$LO or higher, the regulator 
artifacts are typically small. In the $3N$ sector, however, most of 
these artifacts can only be cured at N$^4$LO, when subleading $3N$ contact interactions appear. It is, therefore, important that 
$3N$ interactions are also pushed to sufficiently high orders or other methods are employed to reduce the impact of these regulator artifacts in the $3N$ sector.

As stated above, regulator artifacts reduce in size with
increasing cutoff but observables obtained from chiral interactions are not RG invariant in Weinberg PC. 
In the next section, we will discuss alternative PC schemes that aim to cure the shortcomings of Weinberg PC by restoring RG invariance for physical observables.

\section{Power Counting: Beyond Weinberg}\label{sec:PC}

\noindent
As previously shown, interactions from chiral EFT have been 
successfully employed in calculations of several nuclear
many-body systems. The systematic chiral expansion is based on 
a separation of scales between the typical low momentum scale
$p$ in nuclei and some breakdown scale $\Lambda_b$. The chiral expansion is then constructed by writing 
down the most general Lagrangian consistent with all the symmetries
of low-energy QCD and expanding it in powers of $\{m_{\pi},p\}/\Lambda_b$. In addition, a PC scheme is necessary, which arranges 
interaction operators according to their importance.

Weinberg PC~\cite{Weinberg1979,Weinberg1990,Weinberg:1991um, Weinberg1992}, which is used in the derivation of the majority
of contemporary chiral interactions, is based on naive dimensional
analysis of interaction contributions in terms of momenta. Such a
procedure works well in the pion-nucleon sector due to the 
Goldstone-boson nature of the pions, which allows all amplitudes to be 
expanded in powers of momenta. However, the appearance of bound 
states in the two-nucleon sector complicates this issue and makes 
the nuclear problem nonperturbative. To solve this issue, Weinberg 
suggested to perform a PC in terms of momenta directly 
for the nuclear potential, which includes all irreducible 
diagrams, i.e., diagrams without purely nucleonic intermediate states.
The resulting potential is then used to solve the many-body 
Schr\"odinger or Lippmann-Schwinger equations and, thus, the nuclear 
many-body system, which would generate all the additional 
contributions. The hope in Weinberg PC is that the order-by-order expansion
for the potential directly translates to an order-by-order
expansion for the observable. However, it is not yet clear if that actually is the case.

In general, when building an EFT, it is important that the theory 
show an order-by-order convergence, exhibit a good agreement with
low-energy experimental data, and satisfy RG invariance. Also, a rigorous justification of an EFT truncation 
error hinges on establishing a consistent PC. As was 
stated before, even though Weinberg PC is widely used and
has proven very successful, it has several shortcomings.
First, observables that are obtained are 
not RG invariant, i.e., at each order in the chiral series there are 
counterterms, i.e. LECs, missing that are needed to absorb the residual cutoff 
dependence. In particular, counterterms should appear in all partial 
waves with attractive tensor interactions, where an attractive 
singular interaction with $1/r^n, \,n\ge 3$ behavior 
appears~\cite{Beane:2000wh,Nogga:2005hy}. In Weinberg PC, 
these interactions appear in the form of the one-pion exchange (OPE) already at LO and 
contribute to all partial waves, but there are no counterterms for 
$l>0$ present at this order. In addition to the 
missing RG invariance, for singular attractive interactions, due to
the oscillatory nature of the wave function for $r\to0$, spurious 
bound states can appear, which additionally may lead to problems 
for certain many-body methods that converge to the lowest energy 
states.

Several solutions to this problem have been proposed. Kaplan, 
Savage, and Wise (KSW)~\cite{Kaplan:1998tg, Kaplan:1998we} 
suggested expanding the $NN$ scattering amplitude instead of the
potential, and to treat only LO contact interactions 
nonperturbatively, while all other interaction contributions, i.e., 
other contact interactions and pion exchanges, are treated 
perturbatively. Nevertheless, KSW PC showed a poor 
convergence in the PWA of spin-triplet channels at
N$^2$LO~\cite{Fleming:1999ee} and led to large N$^2$LO corrections. 
It was found that it is necessary to treat pion-exchange diagrams 
nonperturbatively in some spin-triplet channels due to the large 
and singular nature of the OPE tensor force~\cite{Beane:2001bc}.

\begin{figure}[t]
\centering
\includegraphics[trim= 0cm 0.0cm 0.0cm 0.0cm, clip=,width=0.8\columnwidth]{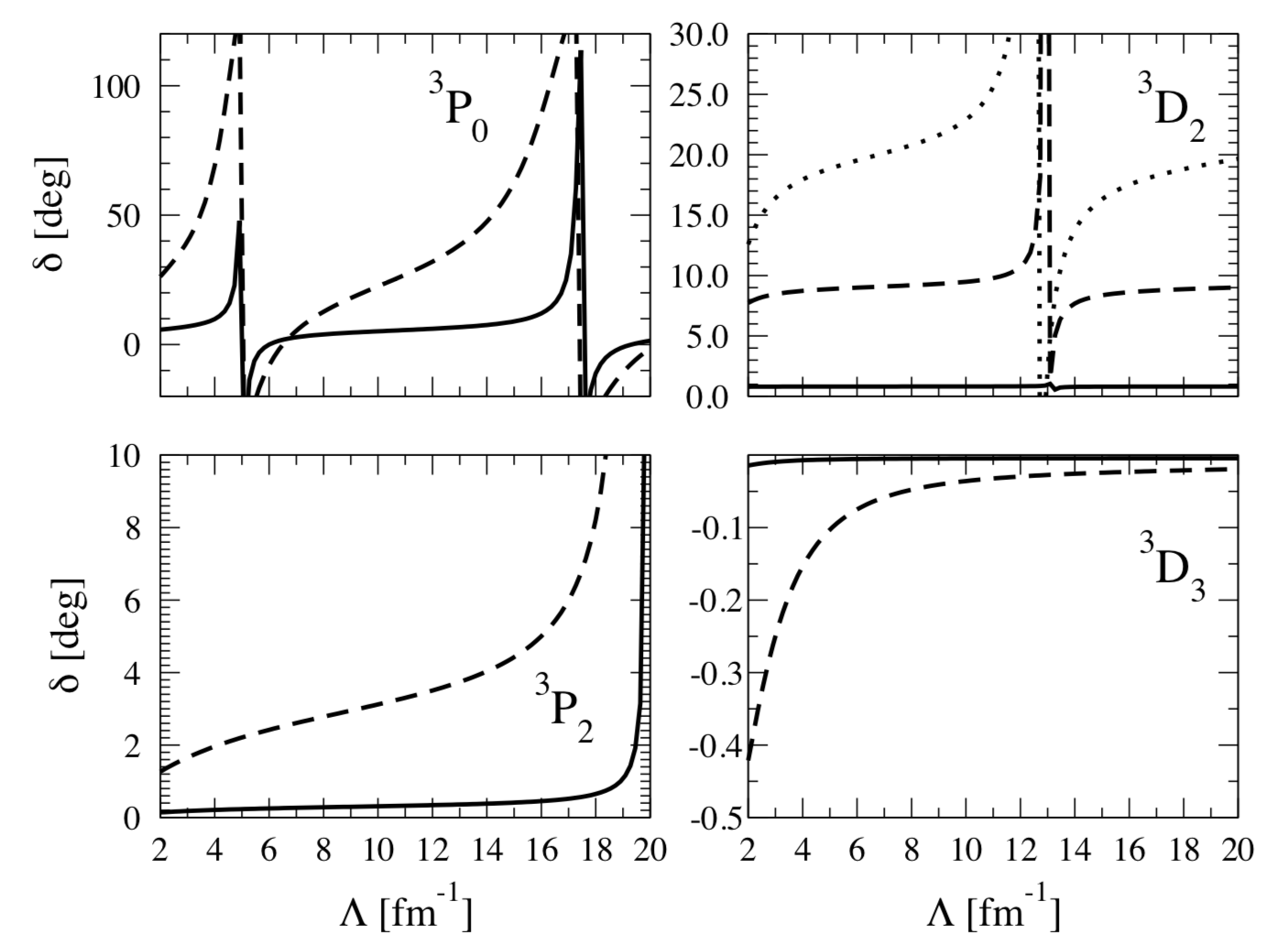}
\caption{\label{fig:NTvK} Cutoff dependence of the spin-triplet
  partial waves $^3P_0$, $^3P_2$, $^3D_2$, and $^3D_3$ with attractive
  tensor contributions at LO in Weinberg PC at laboratory energies of 10 MeV
  (solid line), 50 MeV (dashed line), and 100 MeV (dotted
  line). 
  Reprinted figure with permission from Ref.~\cite{Nogga:2005hy}. Copyright 2005 by the American Physical Society.}
\end{figure} 

Another important contribution was made by Nogga, Timmermans, and 
van Kolck (NTvK)~\cite{Nogga:2005hy}, who studied the cutoff 
dependence of phase shifts at LO in Weinberg PC and 
found strongly cutoff-dependent results as well as spurious bound 
states in partial waves with attractive tensor contributions, see
Fig.~\ref{fig:NTvK}. This cutoff dependence appeared already for
cutoffs of the order of the breakdown scale. To make the interactions RG invariant, NTvK explicitly
added counterterms to the $^3P_0$, $^3P_2$, and $^3D_2$ partial waves,
where tensor interactions are attractive, while higher partial waves
are screened by the centrifugal barrier and therefore remain
perturbative. Similar findings and confirmations of NTvK power
counting were obtained in, e.g., 
Refs.~\cite{Birse:2005um, PavonValderrama:2005wv, PavonValderrama:2005uj}. 
A recent development by Kaplan in performing analytic high-order perturbative calculations with a $1/r^3$ potential shows that only in the ${^3}P_0$ and ${^3}S_1-{^3}D_1$ partial-waves the convergence issue arises, and a possible way is suggested for improvement~\cite{Kaplan:2019znu}.

\begin{figure}[t]
\centering
\includegraphics[trim= 0cm 0.0cm 0.0cm 0.0cm, clip=,width=0.49\columnwidth]{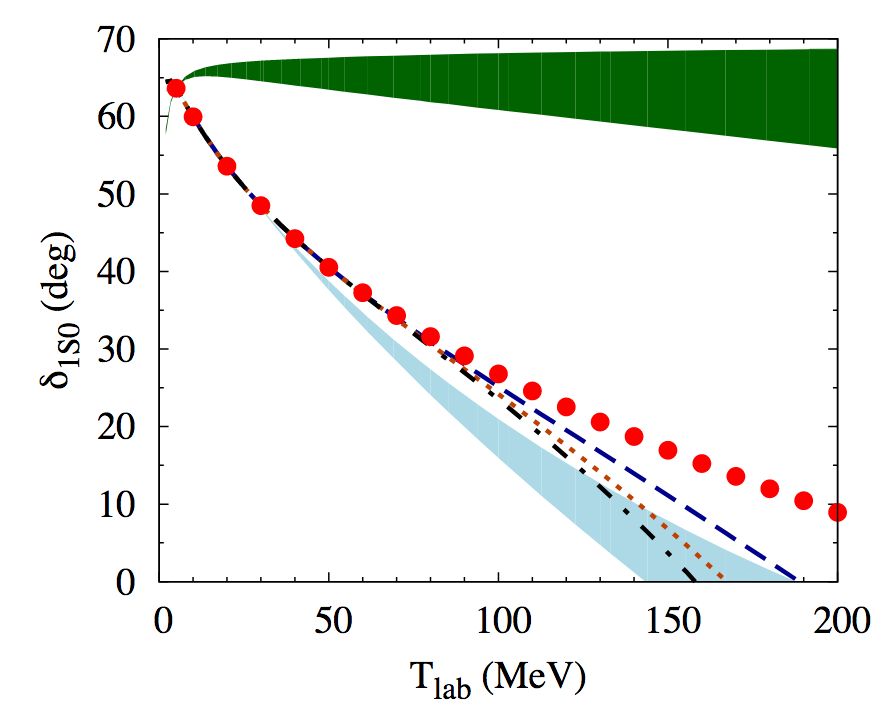}
\includegraphics[trim= 0cm 0.0cm 0.0cm 0.0cm, clip=,width=0.485\columnwidth]{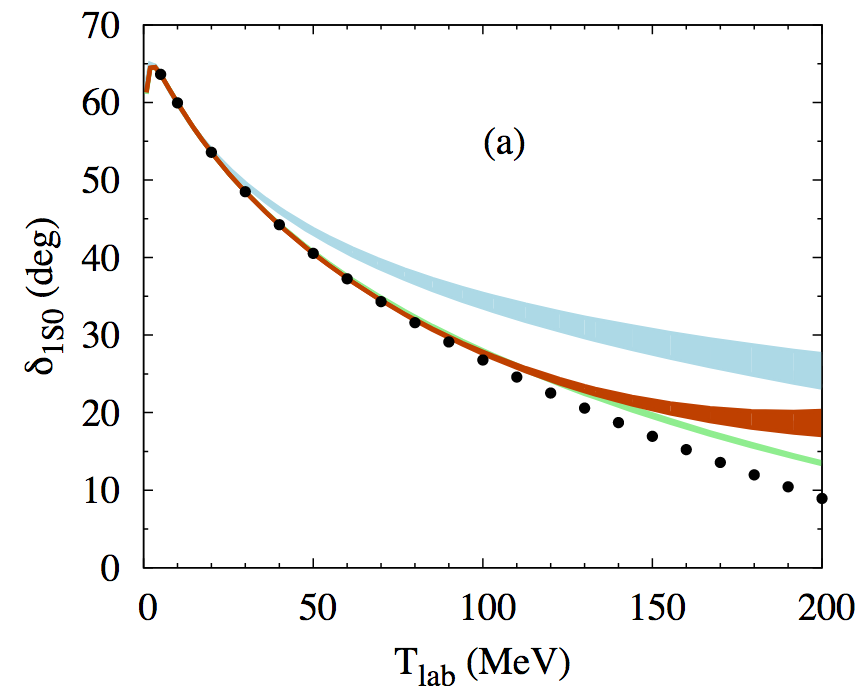}
\caption{\label{fig:LongPC} Nucleon-nucleon scattering phase shifts in the $^1\!S_0$ channel at various orders compared to the Nijmegen partial-wave analysis (red dots in the left panel, black dots in the right panel). In the left panel, the PC of Ref.~\cite{Long:2012ve} is used with the green band at $\mathcal{O}(Q^0)$, the blue band at $\mathcal{O}(Q^1)$, and the dashed, dotted, and dashed-dotted lines at $\mathcal{O}(Q^2)$. In the right panel, the PC of Ref.~\cite{Long:2013cya} is used with the blue band at $\mathcal{O}(Q^{-1})$, the green band at $\mathcal{O}(Q^0)$, and the orange band at $\mathcal{O}(Q^1)$. The bands are obtained by varying the cutoff roughly between $0.5$ and $2$ GeV.
Reprinted figures with permission from Refs.~\cite{Long:2012ve,Long:2013cya}. Copyright 2012 and 2013 by the American Physical Society.
}
\end{figure} 

This strategy, i.e., to require RG invariance of nuclear observables
as the cutoff is taken to values $\Lambda \gg \Lambda_b$, has lead to several new PC schemes. All of these schemes, however, contain 
many features of Weinberg PC. Most of them are constructed by 
testing existing schemes in several partial waves and curing all 
cutoff dependencies or inconsistencies by promoting 
interaction terms, or by treating certain parts of the interaction 
perturbatively or nonperturbatively; see also Refs.~\cite{Valderrama:2009ei,Valderrama:2011mv,Valderrama:2012jv}. 
In Fig.~\ref{fig:LongPC}, for example, we show results for the 
$^1S_0$ phase shift at LO, NLO, and N$^2$LO in the two modified PC schemes of Refs.~\cite{Long:2011qx, Long:2011xw, Long:2012ve, Long:2013cya}. Both schemes lead to RG invariant results in 
the $^1S_0$ channel, but the order-by-order convergence is clearly 
different and might be used to choose one PC scheme over the other. 
The question arises if there is a systematic {\it a priori} way to
construct a PC scheme for which it is not necessary to 
explicitly check its RG behavior in calculations. 

In addition, several problems persist even with modified PC schemes. 
Even though RG invariance is established, in some of the modified
schemes spurious bound states still appear in certain $NN$
scattering channels. As stated before, these may present a 
difficulty for some many-body methods that converge 
to the ground state for a given Hamiltonian, i.e., the state with 
the lowest energy. Compared to the deuteron binding energy of about 
$2$ MeV, spurious bound states can have a much higher binding energy, 
of the order of a few GeV. Some many-body methods, e.g., QMC methods,
would converge to this deep spurious bound state. 

Furthermore, the modified PC schemes typically require 
the interaction at LO be fully iterated. This can not be done in
perturbative many-body methods, e.g., MBPT, and the question arises if perturbation theory works at LO. 
In addition, certain parts of the
short-range interactions might be promoted only in specific partial 
waves and might be treated differently, i.e., perturbative or 
nonperturbative. This may present another difficulty for many-body
methods that are not based on a partial-wave formalism. These
difficulties complicate the practical implementation of modified 
PC schemes in modern many-body methods, but these implementations are necessary to test these schemes in observables besides scattering phase shifts, e.g., binding energies and radii or other static properties of nuclei. 

Finally, the analysis of RG invariance for modified PC 
schemes in heavier nuclear systems might pose a challenge. To explicitly demonstrate RG invariance in these systems, high-cutoff potentials need to be used in calculations. These potentials, however, are too “hard” to be
treated in most of the many-body methods, i.e., they contain a strong
repulsive short-range component, the so-called hard core, that makes
the many-body method converge very slowly if at all. Different
many-body methods will break down at different cutoff scales. 
For instance, it was shown that many-body perturbation theory breaks down at cutoffs of the order of $550$ MeV~\cite{Drischler:2017wtt}.
However, as one does not expect five-nucleon and higher many-body forces to be important in nuclear systems, it may be sufficient 
to establish RG invariance in smaller nuclear systems, 
e.g., $A\leq 5$, and to show that the estimated truncation 
uncertainties are meaningful. Then one can use interactions from these 
PC schemes at a few cutoff scales of the order of the 
breakdown scale in \textit{ab initio} many-body methods. 
But even in this case, a few technical questions remain. Several \textit{ab initio} many-body methods for nuclear structure calculations use the similarity renormalization group~\cite{PhysRevC.75.061001} (SRG) to soften interactions
and it needs to be clarified how the SRG fits into this scheme. 

In any case, in order to implement interactions from improved 
PC schemes in advanced many-body methods and to address 
all of these questions, the EFT community needs to provide 
computational routines to practitioners that deliver, e.g., momentum-space 
matrix elements and already project out all deep spurious bound states.
Given all the effort invested into constructing new PC
schemes, there are already interactions available at LO for several
different schemes. The many-body community should test these
interactions as soon as the corresponding interaction routines are
made available. 
In this context, the participants of the workshop
agreed that it would be extremely useful to develop a general
potential routine that includes all possible interaction
contributions, adds them up according to a chosen PC
scheme, and provides the interaction for this choice. It would be
desirable if such a routine was maintained and kept up-to-date by 
the community to include all available schemes and 
their developments.

Finally, it is worth mentioning that the necessity of adding additional counterterms was debated for higher orders in 
Ref.~\cite{Epelbaum:2006pt}, where it was argued 
that the momentum-space cutoff should be chosen of the order of the breakdown scale and not much larger. 
In particular, it was argued that 
there is no need to use a momentum-space cutoff larger than 
$3~\rm{fm}^{-1}$ and that at this scale the low-energy data shows no inconsistency in Weinberg PC. Using a simple analytically 
solvable model, Ref.~\cite{Epelbaum:2009sd} argued that in an approach similar to the Weinberg PC, taking the cutoff much larger than the breakdown scale is not legitimate from an EFT point of view as it destroys low-energy theorems. Also, when taking the cutoff to infinity, an infinite number of relevant counterterms would have to be included.
Therefore, demanding RG invariance is argued to be a poor criterion for analyzing nuclear EFT consistency. 
This viewpoint is supported by the fact that current chiral 
interactions within Weinberg PC lead to an excellent 
description of nuclear systems also for low cutoffs. However, when 
using Weinberg PC with its inconsistencies, it might be that even though nuclear data close to the valley of stability or $NN$ scattering is properly described, any extrapolation away from the fitting region may not be reliable.

\begin{figure}[t]
\centering
\includegraphics[trim= 0cm 0.0cm 0.0cm 0.0cm, clip=,width=0.8\columnwidth]{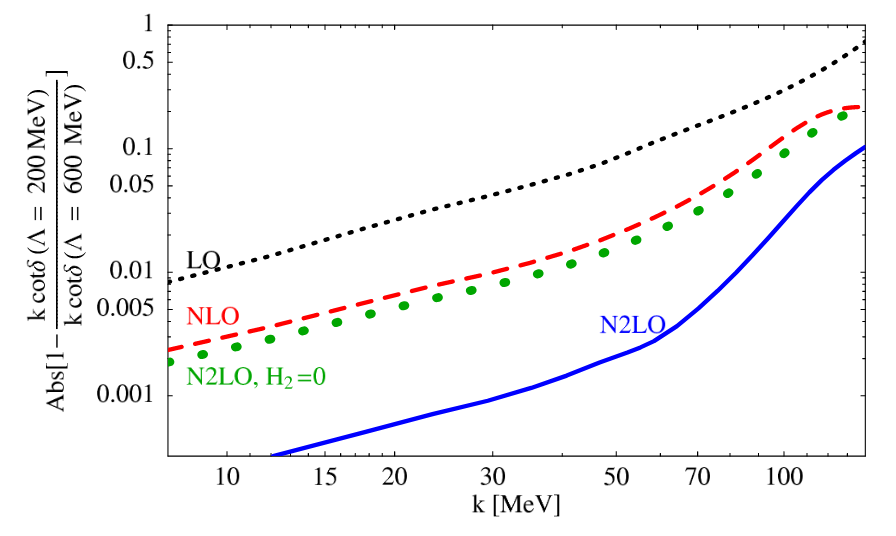}
\caption{\label{fig:Griesshammer}
Double-logarithmic error plot for the $^2S_{1/2}$ partial wave of 
$nd$ scattering in pionless EFT~\cite{Griesshammer:2015osb,Griesshammer:2020fwr}. Figure reprinted from Ref.~\cite{Griesshammer:2015osb} with the permission of H.~Griesshammer.}
\end{figure} 

Independent of this viewpoint, using Weinberg PC offers the pragmatic
possibility to obtain some information about the ``true'' PC. For
this, however, it is imperative to provide calculations order by order
with reproducible estimates for the truncation uncertainties. From
such pragmatic calculations, the quality of the model can then be 
estimated by employing statistical tools discussed above, or other
analysis tools. 
One such tool is a ``Griesshammer'' plot. Any computed 
observable $O(k, \Lambda)$ at a momentum $k$ will carry some residual 
dependence on the cutoff $\Lambda$. One can write $O$ as 
\begin{equation}
O_n(k,\Lambda) =\sum_{i=0}^{n} O_i(k) Q^{i}+C(k,\Lambda)Q^{n+1}\,,
\end{equation}
where the first part is the normalized cutoff-independent part of 
the computed observable and the last term contains the remaining cutoff dependency which is absorbed in some number $C$. Computing $O$ at
order $n$ and at different cutoff scales $\Lambda_1$ and $\Lambda_2$, 
i.e., calculating
\begin{equation}
\frac{O_n(k,\Lambda_1)-O_n(k,\Lambda_2)}{O_n(k,\Lambda_1)}=Q^{n+1} \frac{C(k,\Lambda_1)-C(k,\Lambda_2)}{O_n(k,\Lambda_1)}\,.
\end{equation}

and plotting the result double logarithmically as a function of the momentum $k$ allows one to extract information on the convergence of 
the EFT, even though the cutoff dependency is only a lower limit to 
the ``true'' uncertainty. An example is shown in 
Fig.~\ref{fig:Griesshammer} for the $^2S_{1/2}$ partial wave of $nd$ scattering in pionless EFT. For instance, since $C$ is simply a 
number, the slope of curves in such a plot is determined by the order 
of the calculation. By comparing the calculated slope with the 
predicted one, one can make a statement on the convergence pattern
and consistency of the EFT before making a detailed comparison to 
data. The Griesshammer plots are distinct from Lepage plots, where the 
prediction is compared to data. The advantage of Griesshammer plots
is that the EFT at hand might converge but not to experimental data. 
In such a case, Lepage plots would obscure the convergence whereas 
Griesshammer plots remain useful. Also, the breakdown scale can be 
estimated from such Griesshammer plots. In the given example, one 
can see that the cutoff dependence decreases order by order. 
Also, the fitted slopes agree well with the theoretically predicted 
ones.

Despite all the challenges, it is crucial for the many-body community to 
implement new PC schemes to make progress in
addressing the shortcomings of chiral interactions and the question 
of renormalization in the EFT. The participants of the
workshop agreed that a useful first step would be to implement a 
perturbation around LO, i.e., treat only the LO interaction 
nonperturbatively and to include higher orders in perturbation 
theory, even when working in Weinberg PC. It remains to 
be seen how practical such an approach is in many-body calculations. 

\section{Constraining Nuclear Forces from Lattice QCD}~\label{sec:lqcd}

\noindent
Lattice QCD, a numerical approach to directly solve QCD on a finite space-time
lattice, offers the possibility of
extracting information on systems that are difficult to access
experimentally, e.g., pure neutron systems or systems containing
hyperons. Lattice QCD is computationally expensive, and 
has so far provided results for systems with small mass numbers $A < 5$ and/or 
at large values of the quark masses. However, there has been tremendous progress in lattice-QCD studies of nuclei over the past years, promising a more central role for this program in the near future. Here, we address the question of the prospect of using inputs from lattice QCD to construct nuclear interactions.

Lattice-QCD studies are performed in a finite Euclidean spacetime so that a necessary step in these calculations, the Monte Carlo sampling of QCD gauge-field configurations, becomes plausible. As a result, a direct connection to real-time observables and scattering amplitudes is lost, demanding nontrivial mappings between the output of lattice QCD and interesting observables in the few-nucleon sector. 
The spectral decomposition of Euclidean correlation functions calculated with lattice QCD gives access to hadron masses and binding energies of deeply bound multi-hadron states directly, up to exponentially suppressed corrections in volume that scale like $e^{-L/R}$, where $L$ is the size of the lattice.
Here, $R$ denotes the finite range of interactions, either the Compton wavelength of the pion, or the inverse binding momentum of the bound state. 
To access scattering amplitudes and transition rates, however, finite-volume energy eigenvalues and matrix elements extracted from lattice QCD must be used to indirectly construct the desired quantities. In the two-hadron sector, L\"uscher's formalism~\cite{Luscher:1986pf, Luscher:1991n1} and its generalizations~\cite{Rummukainen:1995vs, Kim:2005gf, He:2005ey, Davoudi:2011md, Hansen:2012tf, Briceno:2012yi, Briceno:2013lba,Gockeler:2012yj}, provide such a mapping rigorously, up to exponentially suppressed corrections in volume. Its success when applied to studies of (single or coupled) two-hadron scattering and resonances is notable, see e.g.,~\cite{Wilson:2015dqa, Briceno:2016mjc, Brett:2018jqw, Guo:2018zss,  Andersen:2018mau,  Dudek:2016cru, Woss:2019hse, Orginos:2015aya, Berkowitz:2015eaa, Wagman:2017tmp}, and has sparked massive analytical and numerical results in recent years in order to extend the mapping to inelastic transition amplitudes involving two hadrons~\cite{Lellouch:2000, Meyer:2011um, Briceno:2012yi, Bernard:2012bi, Briceno:2014uqa, Feng:2014gba, Briceno:2015csa, Briceno:2015tza}, as well as elastic scattering of three-hadron systems~\cite{Polejaeva:2012ut,Briceno:2012rv, Hansen:2014eka,Hansen:2015zga, Hammer:2017uqm,Hammer:2017kms, Guo:2017ism,Mai:2017bge, Briceno:2017tce, Mai:2018djl, Blanton:2019igq,Blanton:2019vdk,Mai:2019fba}.

In the nuclear sector, the first calculation of lowest-lying spectra of light nuclei and hypernuclei up to $A<5$ was reported in Ref.~\cite{Beane:2012vq}, and since then a great deal of effort has been devoted to constraining low-energy scattering parameters of two-baryon systems directly from lattice QCD via applications of L\"uscher's method, albeit still at unphysically large values of the quark masses. 
Ideas for improving finite-volume effects in binding energies of shallow nuclear bound states have been put forward~\cite{Davoudi:2011md,Briceno:2013bda,Briceno:2013hya}, and formalisms that allow extractions of not only the $S$-wave but also higher partial-wave amplitudes from lattice QCD have been developed~\cite{Briceno:2013lba} and successfully implemented~\cite{Berkowitz:2015eaa}. Studies at lower quark masses are underway, and will be becoming more sophisticated with increased computational resources, as one moves towards addressing issues such as signal-to-noise degradation~\cite{Wagman:2016bam}, excited-state contamination, and energy ``plateau'' identification in nuclear correlation functions.

To connect these studies to the few- and many-body sector of nuclear physics, a clear path forward is to match EFT interactions to the corresponding lattice-QCD results.~\footnote{See Ref.~\cite{Drischler:2019xuo} and references therein for a proposal based on matching lattice QCD to the ``Harmonic Oscillator Based Effective Theory''}. A matching of the complete 
set of LECs to lattice-QCD data provides the possibility of computing 
results truly from first principles, but certain pieces of 
experimental data will remain much more precise than results from lattice QCD. However, for systems where experimental
information is scarce or does not exist, a partial matching of 
nuclear interactions to lattice-QCD data may be the most practical approach. In fact, instances of such matching have emerged in the nucleon and hyperon sector in recent years, including predictions for the significance of $\Sigma$ hyperons in the decomposition of the interior of neutron stars obtained from lattice-QCD calculations of $N\Sigma$ scattering~\cite{Beane:2012ey}, estimations of the binding energy of the long-sought-for H-dibaryon~\cite{Beane:2010hg,Beane:2011zpa,Inoue:2010es}, evidence for spin-flavor symmetry and an $SU(16)$ extended symmetry in the $B=2$ octet-baryon sector of QCD observed at $m_\pi \approx 800$ MeV~\cite{Wagman:2017tmp} in accordance with large-$N_c$ predictions~\cite{Kaplan:1995yg}. 
These conclusions were only possible using lattice QCD given the scarcity of constraints on hypernuclear interactions in nature.

\begin{figure}[t]
\centering
\includegraphics[trim= 0cm 0.0cm 0.0cm 0.0cm, clip=,width=0.8\columnwidth]{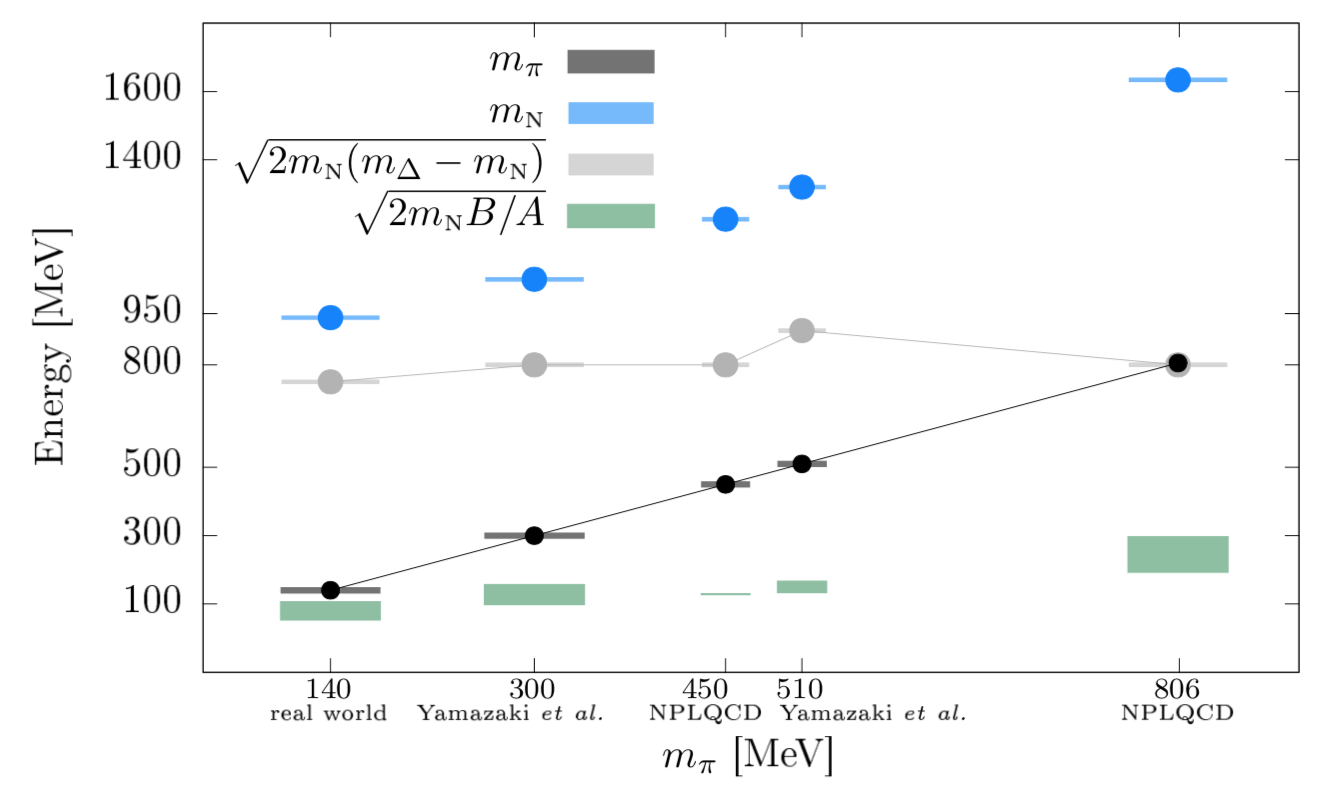}
\caption{\label{fig:sepscalesmpi}
Separation of scales for nuclear EFTs as a function of pion mass. 
Reprinted figure with permission from Ref.~\cite{Kirscher:2015tka}. Copyright World Scientific Publishing Company.}
\end{figure}

\begin{figure}[t]
\centering
\includegraphics[scale=0.6]{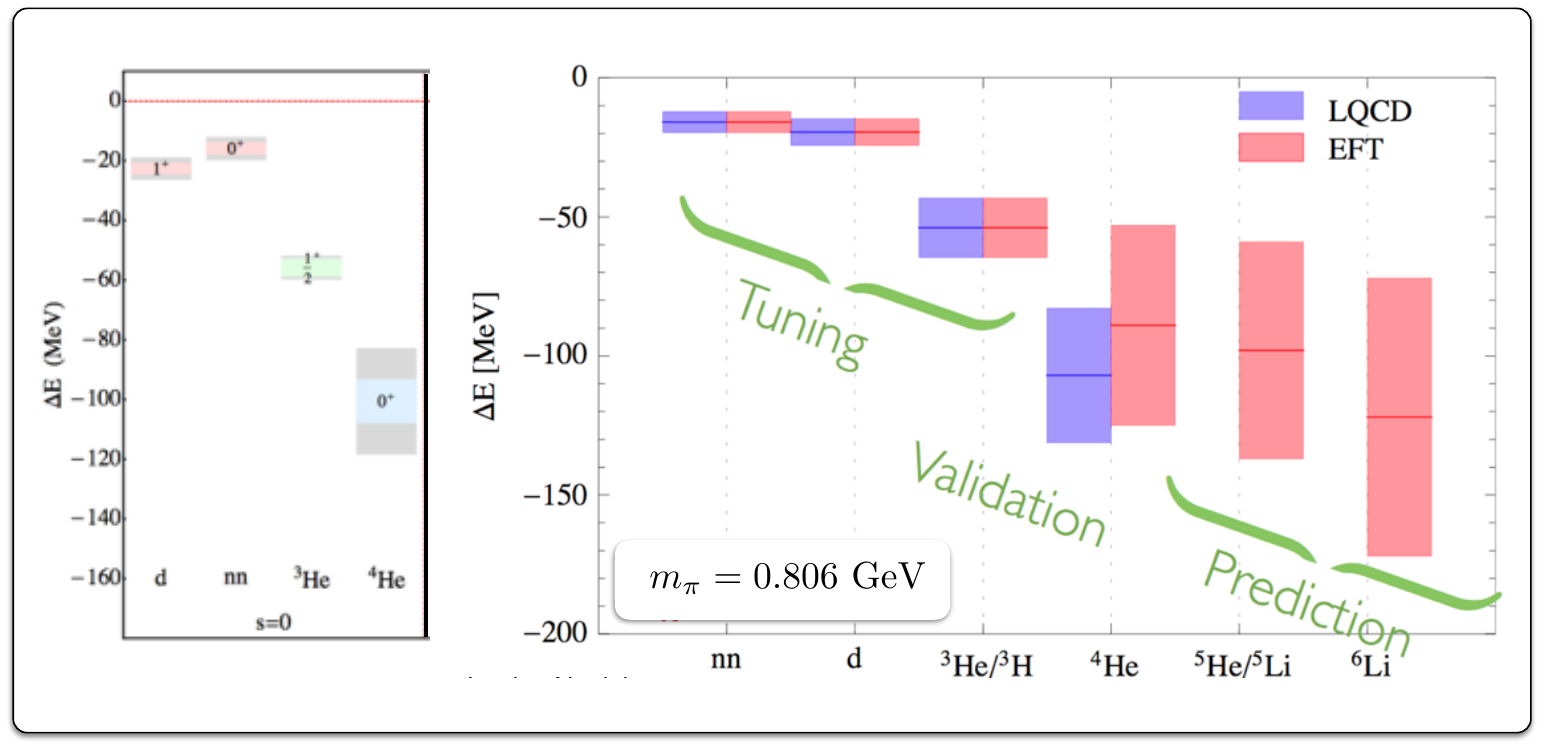}
\caption{\label{fig:matching800mev}
[Left panel] Binding energies of the deuteron, dineutron, $^3$He and $^4$He at an SU(3) flavor-symmetric point obtained with lattice QCD at a pion mass of $\sim 800$ MeV. [Right panel] The binding energies of light nuclei up to $A=6$ obtained from a QMC computation with lattice-QCD input ($A=2,3$ binding energies in the left) for the LECs of the pionless EFT. The left plot is reproduced from Ref.~\cite{Beane:2012vq} with permission of the NPLQCD collaboration and the American Physical Society. Copyright 2013 by the American Physical Society. 
The right plot is a compilation of the results of Refs.\cite{Beane:2012vq, Barnea:2013uqa,Kirscher:2015tka}, reproduced from Ref.~\cite{Savage:2016egr} with the permission of M.~Savage.}
\end{figure}

Another example of such a matching appeared in a series 
of works where the LECs of pionless EFT were fit to lattice-QCD 
results at several pion masses~\cite{Barnea:2013uqa,Kirscher:2015yda, Contessi:2017rww}. This approach works very well at large pion masses 
because the separation of scales is stronger and pionless EFT may work 
better, see Fig.~\ref{fig:sepscalesmpi}.
In Ref.~\cite{Contessi:2017rww}, pionless EFT interactions were
constructed at LO for several pion masses ($140$~MeV, 
$510$~MeV, and $805$~MeV) and various cutoff scales and then tested 
in nuclear-structure calculations using QMC methods. 
For \isotope[4]{He}, the results at the physical pion mass were very close 
to experiment, and at the larger pion masses, in very good agreement with the
lattice-QCD data, see Fig.~\ref{fig:matching800mev}. However, it was found in this approach that the 
\isotope[16]{O} nucleus is unbound with respect to break-up into 
four $\alpha$ particles at LO for all quark masses. In addition, it
was argued that including NLO corrections perturbatively will 
not resolve this problem as both the binding energy of
\isotope[16]{O} and the four-$\alpha$ threshold move in the same 
direction. When including NLO corrections nonperturbatively,
however, \isotope[16]{O} was found to be bound~\cite{Bansal:2017pwn}. 
Thus, the existence of a bound \isotope[16]{O} would only be possible at N$^2$LO in a proper PC. This, in turn, leads to the question of whether pionless EFT is the correct approach to address such a large-$A$ system, and matching a chiral EFT with explicit pion-exchange contributions to the lattice-QCD data may prove to be more successful.

Another intrinsically different alternative to L\"uscher's method is advocated and practiced by the HAL-QCD
collaboration~\cite{Ishii:2006ec,Aoki:2008hh,Aoki:2011gt,HALQCD:2012aa}. In this approach, the
Nambu-Bethe-Salpeter (NBS) wave function $\psi^{\alpha}(E,\textbf{x})$
is extracted from the baryon four-point correlation function calculated using lattice QCD. An energy-independent but nonlocal potential
$U(\textbf{r},\textbf{r}')$ is then defined such that the NBS wave function
obeys a Schr\"odinger-like equation in finite volume,
\begin{equation}
(p^2+\nabla^2) \psi^{\alpha}(E,\textbf{x})=\int d^3y U_{\beta}^{\alpha}(\textbf{x},\textbf{y})\psi^{\beta}(E,\textbf{y})\,.
\end{equation}
A local potential $V(r)$ is then extracted by assuming a truncated derivative expansion of 
$U$, and is required to reproduce the data from the corresponding lattice-QCD calculation. 
This method has been employed to obtain the two- and three-nucleon potentials, which are then used in infinite volume to solve the Lippmann-Schwinger equation for two-baryon scattering phase shifts away from and recently at the physical quark masses~\cite{Ishii:2006ec,Aoki:2008hh,Aoki:2011gt,HALQCD:2012aa,Gongyo:2017fjb,Miyamoto:2017tjs,Iritani:2018sra}. The resulting nuclear potentials have also been used in nuclear-strcuture calculations, for example for doubly-magic nuclei using the SCGF method for a pion mass of $m_{\pi}=496$ MeV~\cite{McIlroy:2017ssf}.

The primary question raised is how reliable these nuclear potentials are, given that the potential is not a physical observable, and is dependent upon the choice of interpolating lattice-QCD operators used to construct correlation functions. Energies on the other hand, are physical observables that can, in principle, be extracted from lattice-QCD correlators, apart from practical issues of reliability of the assumption of ground-state saturation due to finite statistics. While it is known that the result of scattering phase shifts using HAL-QCD method should agree at the energy eigenvalues of the volume with those obtained from L\"uscher's formula~\cite{Kurth:2013tua,Beane:2010em,Iritani:2018vfn}, it is not known how close one is to the correct values away from these eigenvalues. While the quoted benefit of (new variants of~\cite{HALQCD:2012aa}) the HAL-QCD method is to avoid the difficulty of energy-``plateau'' identification~\cite{Iritani:2017rlk,Beane:2017edf,Davoudi:2017ddj,Yamazaki:2017jfh,Gongyo:2017fjb}, further studies must be conducted to estimate systematic uncertainties associated with the use of scheme-dependent NBS wavefunctions and subsequently potentials in a finite discretized spacetime, and with a finite-order derivative expansion for the nonlocal potential~\cite{Beane:2010em,Yamazaki:2017gjl,Iritani:2018zbt}. The direct use of these potentials in many-body nuclear calculations is, in particular, questionable given the uncontrolled uncertainty in the potential, especially at short and medium length scales that may be more relevant in dense systems of nucleons.

In order to follow the path of matching nuclear forces to QCD, one must explore the opportunities that are more practical for the time being, given the computational cost of lattice-QCD studies of nuclei, and the formal complexity of the mapping of scattering and reaction observables to Euclidean finite-volume correlation functions as the number of nucleons grows. For example, one may bypass L\"uscher's methodology and its developing generalizations, to directly match energy eigenvalues to those obtained with EFT interactions using many-body techniques that are applied in finite volume with the same boundary conditions, hence fixing the EFT LECs at given a renormalization scale. The first steps along these lines have already been taken, see e.g., Refs.~\cite{Beane:2012ey, Klos:2016fdb, Contessi:2017rww}. It is conceivable that this approach in the three-neutron system will have the most immediate impact. This, however, requires precise three-neutron spectra at low energies to be produced by the lattice-QCD community at or near the physical quark masses, which appears to be challenging in the near term due to its computational cost.

Another conceptually and practically interesting question is whether constraints on nuclear forces at unphysical values of the quark masses can teach us something new about the nature of these forces in nature. Some examples are: (1) How valuable will constraints at near-physical quark masses be in constraining the LECs of chiral EFTs, and how useful will disentangling the pion-mass dependence of contact interactions be at each order in the EFT in making predictions for new observables? (2) Can lattice-QCD studies at larger quark masses shed light on power-counting issues given the different separation of scales in play, and offer some lessons for the physical world? (3) Do nuclear systems at low energies retain their unnatural features, such as near-threshold bound states, large scattering lengths, apparent fine-tuning in the existence of the Hoyle state and in the carbon production rate, etc., even if the masses of the quarks were set drastically differently~\cite{Beane:2002xf,Bedaque:2010hr,Epelbaum:2013wla,Epelbaum:2012iu}? In other words, are these features robust, and solely a consequence of Yang-Mills theory with $N_c=3$, regardless of the value of the constant parameters in the Lagrangian? Lattice QCD appears to be on the path to addressing, directly or indirectly, some of these fundamental questions~\cite{Beane:2013br,Yamazaki:2015asa}, which can be exciting if successful. However, before making any final conclusion on these questions, it would be important that different nuclear lattice-QCD collaborations resolve discrepancies~\cite{Davoudi:2017ddj} and converge on a set of results that can be considered highly reliable.

\section{Conclusions and Outlook}
\noindent
This review summarizes the discussions that took place, and the conclusions that were reached, at a 2018 ECT* workshop on ``New Ideas in
Constraining Nuclear Forces''. It contains an analysis of the current limitations of
nuclear forces, which present a major bottleneck towards accurate theoretical predictions of properties of nuclei. The  main sources of uncertainty are the
truncation of the EFT expansions of forces, the missing RG invariance in popular approaches, and hence
the dependence on the regularization scheme and scale. These lead to inaccurate and/or imprecise few- and many-nucleon forces, which add to additional uncertainties associated with the many-body methods when applied to certain observables. The main conclusions outlined in this review can be summarized as follows:
\begin{itemize}
\item[$\square$]{Nuclear forces can be constrained using an improved
set of experimental observables. In fact, it would be extremely valuable to identify
a set of ``golden'' observables that capture many important features of nuclear forces, and that all nuclear interactions and many-body methods should reproduce. These observables should be adopted by the
nuclear-physics community, and be chosen, and demonstrated to be, complementary. For this
set of observables, it would be ideal if the community maintains a
database that collects results for different methods and interactions in order to gauge their relative reliability, as emphasized below.}

\item[$\square$]{Nuclear forces can be improved with novel fitting
strategies and the inclusion of higher orders in the EFT power counting. There has been tremendous progress in
pushing chiral interactions to higher orders and constraining them
using, e.g., Bayesian statistical analyses. This progress highlights an understanding reached by the community that results are only meaningful with reproducible systematic
uncertainty estimates. Therefore, it would be a great step forward if all researchers in this area publish results at several chiral orders, and with the corresponding
estimates for the truncation error.}

\item[$\square$]{The shortcomings of the Weinberg PC can potentially be cured by implementing improved PC schemes in nuclear many-body methods. There has been significant progress in this field since the early era of development of nuclear EFTs. However, as one moves towards realistic implementation of such schemes in larger systems, it will be important that the PC community provide interaction routines that project out spurious bound states, and are available at several cutoff scales, from soft to hard interactions.}

\item[$\square$]{Lattice QCD has the promise of providing useful constraints on observables that are difficult
or impossible to access experimentally. These may lead to constraints on multi-neutron and hypernuclear forces, as well as short-distance contributions to nuclear responses to external probes, such as electroweak and beyond-the-Standard-Model nuclear transitions. Lattice QCD is currently unable to provide results that are
competitive with experimental data in the few-nucleon sector. However, the pace of progress suggests that the method may become an alternative, or in some instances the primary, tool in constraining observables, and a complementary source of input to refine nuclear forces, in the upcoming decade.}
\end{itemize}

As a final word, and to further emphasize the role a community-wide coordinated effort can play in successfully constraining and utilizing EFT-based nuclear forces and currents, we propose the creation of a database for the theory results for all phenomenologically important nuclear observables. The database should compile the associated results, as well as report on strengths and shortcomings of a given study. Such an effort has a long-established and successful presence in the lattice-QCD community. The Flavor-Lattice Averaging Group (FLAG), comprised of 
representatives of all relevant international collaborations, regularly publishes a compilation of a number of lattice-QCD results~\cite{Aoki:2019cca}, i.e., those that serve for verification purposes, or for informing the corresponding experimental programs, along with an agreed-upon qualification system that rates the rigor of each result. It, therefore, provides a resource to the wider community, and shows to what extent each number can be trusted. With the growth of nuclear few- and many-body calculations, and the diversity of inputs, methods, and analysis approaches, it may be worth trying to organize systematically a similar program in the low-energy nuclear structure and reaction community. This may bring further synergy to the field, shed light on the underlying problems, and lead to progress by providing more coherence and transparency.

\section*{Acknowledgments}
\noindent
We thank the ECT* for hosting the workshop ``New Ideas in Constraining Nuclear Forces'' and all its participants for excellent contributions
and interesting discussions that have led to this review. We
especially thank Richard J. Furnstahl for inviting us to write this
review. 
This work was supported by the US Department of Energy, Office of 
Science, Office of Nuclear Physics, under Contract DE-AC52-06NA25396, 
the Los Alamos National Laboratory (LANL) LDRD program, the NUCLEI 
SciDAC  program, the European Research Council (ERC) under the European Unions Horizon 2020 research and innovation programme (Grant agreement No. 758027), the Alfred P. Sloan Foundation, the Maryland Center for Fundamental Physics at the University of Maryland, College Park, and by the ERC Grant No.~307986
STRONGINT and the BMBF under Contract No.~05P18RDFN1.

\section*{References}

\bibliographystyle{iopart-num}
\bibliography{toprev}

\end{document}